\documentclass{article}

\usepackage{PRIMEarxiv}

\usepackage[utf8]{inputenc} 
\usepackage[T1]{fontenc}    
\usepackage{hyperref}       
\usepackage{url}            
\usepackage{booktabs}       
\usepackage{amsfonts}       
\usepackage{nicefrac}       
\usepackage{microtype}      
\usepackage{lipsum}
\usepackage{fancyhdr}       
\usepackage{graphicx}       
\graphicspath{{media/}}     
\usepackage{amsmath}

\title{Artificial intelligence pathways from weather to climate}

\author{
\textbf{Tom~Beucler}$^{1\ast}$,
\textbf{J. David~Neelin}$^{2}$,
\textbf{Hui~Su}$^{3}$,\\
\textbf{Shivanshi~Asthana}$^{1}$,
\textbf{Chris~Bretherton}$^{4}$,
\textbf{Will~Chapman}$^{5}$,
\textbf{Costa~Christopoulos}$^{6}$,
\textbf{Spencer~K.~Clark}$^{4,7}$,\\
\textbf{Aditya~Grover}$^{2}$,
\textbf{Ignacio~Lopez{-}Gomez}$^{8}$,
\textbf{Tapio~Schneider}$^{6,8}$,
\textbf{Adam~Subel}$^{9}$,
\textbf{Oliver~Watt{-}Meyer}$^{4}$
\\[1ex]
$^{1}$University of Lausanne, Lausanne, Switzerland.\\
$^{2}$University of California, Los Angeles, Los Angeles, CA, USA.\\
$^{3}$Hong Kong University of Science and Technology, Hong Kong SAR, China.\\
$^{4}$Allen Institute for AI, Seattle, WA, USA.\\
$^{5}$University of Colorado Boulder, Boulder, CO, USA.\\
$^{6}$California Institute of Technology, Pasadena, CA, USA.\\
$^{7}$NOAA/Geophysical Fluid Dynamics Laboratory, Princeton, NJ, USA.\\
$^{8}$Google Research, Mountain View, CA, USA.\\
$^{9}$New York University, New York, NY, USA.\\
$^{\ast}$Corresponding author: \texttt{tom.beucler@unil.ch}
}

\begin{document}
\maketitle

\begin{abstract}
Deep learning has made rapid advances in weather forecasting: autoregressive models trained on atmospheric reanalyses now rival dynamical models across nowcasting, medium-range, and subseasonal-to-seasonal lead times, producing well-calibrated ensemble forecasts at reduced cost. We review these advances and consider their extension to climate horizons, where the challenge shifts from initial-condition skill to producing reliable statistical responses under altered forcings. AI-powered climate prediction systems must produce credible forced responses to drivers (e.g., greenhouse gases, land-use change) typically outside the observed record. We propose two minimum requirements for AI in climate modeling: (i) external forcing agents must enter explicitly enough to support interventions in which they vary independently; and (ii) robustness must be stress-tested in out-of-distribution regimes, including extremes and counterfactual trajectories. Using leading AI autoregressive emulators and hybrid physics--AI models, we identify development and coupling challenges. Comparing the reported throughput of these models with that of GPU-ported dynamical models highlights how AI can reduce time-to-solution by advancing only the target variables at the required resolution and using longer time steps, rather than integrating a full high-frequency, multivariate state. Diverse AI downscaling strategies can partially substitute for explicit fine-scale resolution, paving the way toward inexpensive local hazard assessment across prediction horizons.
\end{abstract}

\section{Introduction}

Earth system prediction is being transformed by artificial intelligence (AI), most clearly in weather forecasting. Following successes in nowcasting \cite{ravuri2021skilful,zhang2023skilful} and medium-range (up to $\sim$2 weeks) forecasting \cite{price2025probabilistic}, data-driven models have recently begun rivaling dynamical (``physics-based'') models on subseasonal lead times of $\sim$1 month when benchmarked using probabilistic scoring \cite{chantry2025son,schloer2026aifs}. They produce forecasts at a fraction of the computational cost \cite{bi2023accurate}, demonstrating that data-driven modeling advances can drive forecast skill, especially for problems that can be framed as a learnable mapping from a known initial state to a known future state.

This progress should not be mistaken for independence from dynamical modeling. Currently, the success of top-scoring data-driven models heavily relies on dynamical models because (i) meteorological reanalyses are used for training and evaluation \cite{rasp2024WB2}, (ii) initial conditions are generally provided by assimilating data into dynamical models \cite{lang2024aifsecmwfsdatadriven}, and (iii) dynamical forecasts are often used as inputs on longer (subseasonal) timescales \cite{mouatadid2023adaptive}. Even end-to-end observation-driven efforts \cite{allen2025end,alexe2024graphdopskilfuldatadrivenmediumrange,sun_data-to-forecast_2025} rely on observational products that embed physical modeling. If dynamical modeling deteriorates, so do the data, initial conditions, and benchmarks on which AI weather models rely. AI therefore shifts, rather than removes, the need for dynamical modeling.

This dependence is paired with a fundamental limitation: AI weather models can underperform dynamical models for unprecedented, record-breaking extremes not seen during training \cite{sun2025canAImodelsOOD,zhang2025numericalmodelsoutperformai,rackow2024robustnessaibasedweatherforecasts}. This out-of-distribution (OOD) forecasting challenge parallels the larger challenge at the heart of climate modeling. Climate projection is not simply weather forecasting extended to longer lead times. Its goal is not to predict a specific day’s weather decades ahead but to estimate how Earth system statistics change under altered boundary conditions, including changing greenhouse gases, aerosols, and land use. Yet because data for many such conditions do not yet exist, and some counterfactual conditions will never be observed, learning forced responses from observations alone appears infeasible. The gap between weather and climate is therefore wider than a difference in lead time, and wider than a simple contrast between initial- and boundary-condition prediction.

We therefore ask what it takes for AI to provide accurate, well-calibrated probabilistic predictions that (i) include rare, potentially OOD events at weather timescales; and (ii) produce physically consistent counterfactual responses under unobserved forcing at climate timescales. We follow these challenges from weather to climate: how to recover event-scale spatial structure and extremes from models optimized for stable forecasts, how successful approaches and predictability evolve as prediction timescales increase, and which climate modeling and evaluation choices are required for credible forced responses. To provide concrete examples, we illustrate these steps with models the authors have helped develop, including design choices and failure modes. Throughout, we distinguish several roles that AI plays as we move across timescales: as a weather prediction system that advances Earth system states forward in time; as a component of a climate prediction system that projects Earth system responses to hitherto unobserved forcings; and as a downscaling tool that renders climate information locally actionable—each placing different demands on the underlying model.

\section{Data-driven forecasting with AI}

AI weather forecasting can be viewed as a family of conditional prediction problems: Given atmospheric information available at and before time $t$, these problems can be organized around a conditional distribution $p(x_{t+\Delta t}\mid x_t, x_{t-\Delta t},\ldots)$ of the future state given current and past information, from which a model may extract different objects: its central tendency (a single future state), samples (an ensemble of plausible futures), a conditional distribution for a derived local observable, or its gradient with respect to the inputs (a forecast sensitivity). These choices define different forecast objects, losses, sampling procedures, rollout strategies, and evaluation criteria, which strongly shape what the model learns, sometimes more so than the architecture alone \cite{garg2026recipe}. 

\subsection{Deterministic AI forecasting}

Deterministic AI forecasting learns a map from the current climate-system state $x_t$ at time $t$, and often one or more past states, to a single future state. This map approximates a summary statistic of the conditional distribution above (in practice, usually its central tendency). In most autoregressive systems, the model is applied repeatedly with a fixed forecast stride $\Delta t$; after $n$ steps of equal strides, the lead time is $\tau=n\Delta t$:
\begin{equation}
\hat{x}_{t+\Delta t}=x_t+f_{\theta}\left(x_t,x_{t-\Delta t},\ldots\right),
\label{eq:deterministic_forecast}
\end{equation}
where $\hat{x}$ denotes a model prediction, $f_{\theta}$ is the learned increment or residual map, and $\theta$ denotes trainable parameters. The residual form emphasizes that many deterministic AI weather models can be viewed as learning state increments rather than absolute states. Their achievement is to learn an autoregressive map that can be iterated to evolve a multivariate climate-system state over lead times of several days to weeks, making deterministic AI weather models predictive and more than statistical generators of meteorological reanalyses.

AIFS \cite{lang2024aifsecmwfsdatadriven}, GraphCast \cite{lam2023GraphCast}, and Pangu-Weather \cite{bi2023accurate} showed that neural networks trained on meteorological analyses or reanalyses can produce stable, skillful multi-day forecasts at global scale. These systems differ substantially in their architectures---graph neural networks, Earth-specific transformers, or hybrid graph-transformer designs---yet recent design studies suggest that no single architectural family is solely responsible for forecast skill once models are expressive enough and the training pipeline is well tuned \cite{siddiqui2024exploringdesignspacedeeplearningbased}. Architectures are therefore less and less of a bottleneck, and frameworks such as Anemoi \cite{prietoNemesio2025Anemoi} that make graph-based architectures easier to implement are helping the field transition from model design towards operationalization through reproducible training and evaluation pipelines.

The success of these models is closely tied to their optimization objective. Most deterministic systems are trained to minimize pointwise errors in future atmospheric fields. Such losses reward forecasts that place the broad circulation, fronts, and synoptic structures in approximately the right location, which is essential for medium-range skill. However, for mean-squared or mean-absolute error losses, the optimal point forecast is a conditional central tendency: the conditional mean for MSE, and the conditional median for MAE. When several sharp futures are plausible, this favors smoother forecasts, since a displaced sharp feature is penalized both where it is missing and where it is falsely predicted \cite{subich2025fixing}. This smoothing can be useful for stable forecasts, but it can also suppress small-scale variance, extremes, and event-scale structure.

In addition to the loss function, forecast stride, rollout training, filtering, and multi-step fine-tuning all interact with the chaotic, scale-dependent growth of forecast error. A 6--12~h stride usually prevents models from predicting fast modes directly, while multi-step training discourages trajectories that leave the data manifold after a few autoregressive steps. These choices help stabilize longer rollouts, but they can also favor low wavenumbers and under-represent grid-scale variance \cite{brenowitz2024LEF,chattopadhyay2024challengeslearningmultiscaledynamics}. Many successful deterministic pipelines therefore behave, explicitly or implicitly, like low-pass filters: AIFS uses down-/up-sampling operations \cite{lang2026aifscrps}, FourCastNet uses spectral truncation \cite{pathak2022fourcastnetglobaldatadrivenhighresolution}, and longer forecast strides can regularize learning in chaotic systems \cite{parthipan2025regularizationmlmodelsearth}. This trade-off is not necessarily a defect as it is one reason deterministic systems can achieve high large-scale skill and stable rollouts, with recent studies suggesting that within AI models, predictability beyond two weeks can arise when initial conditions are optimized \cite{vonich2024predictabilitylimit2021HW,vonich2025testinglimitatmosphericpredictability}. But it certainly motivates probabilistic forecasting and targeted post-processing when the goal is risk estimation or local impact assessment.

\subsection{Probabilistic AI forecasting}

A single deterministic trajectory, being a central tendency of the conditional distribution, cannot represent the range of plausible future atmospheric evolutions. Probabilistic AI forecasting instead targets samples from that same conditional distribution, now learned explicitly as $p_{\theta}$: the two approaches condition on identical information and differ only in the forecast object: a point estimate versus a set of samples. 
Probabilistic AI forecasting learns a conditional distribution $p_{\theta} $,
\begin{equation}
\hat{x}_{t+\Delta t}^{(i)}\sim p_{\theta}\left(\cdot\mid x_t,x_{t-\Delta t},\ldots\right),
\qquad i=1,\ldots,N_e ,
\label{eq:probabilistic_forecast}
\end{equation}
from which the members $\hat{x}_{t+\Delta t}^{(i)} $ can be sampled, where $i$ indexes ensemble members. 

All members share the same initial state; for direct-generation models they may differ through a per-member random input $\xi^{(i)}$ to represent aleatoric uncertainty, or through their generating model $G_{\theta}^{(i)}$ to represent epistemic (model) uncertainty, i.e.\ $\hat{x}_{t+\Delta t}^{(i)}=G_{\theta}^{(i)}\!\left(x_t,x_{t-\Delta t},\ldots,\xi^{(i)}\right)$. This makes AI forecasting closer to operational ensemble prediction: the forecast object is not simply a best estimate, but a collection of plausible trajectories whose spread provides an estimate of forecast uncertainty.

Proper scores such as the Continuous Ranked Probability Score (CRPS) avoid the central-tendency collapse of MSE/MAE losses, but they raise the challenge of multivariate calibration. Recent state-of-the-art systems differ along two axes: the objective used to train the distribution, and the mechanism used to sample from it. A first family uses fast direct ensemble generation. AIFS-CRPS \cite{lang2026aifscrps} and FGN \cite{alet2025skillfuljointprobabilisticweather} map the atmospheric state, together with random input used to distinguish ensemble members, directly to samples from a forecast distribution. These approaches are attractive because inference is comparatively cheap: once trained, producing additional ensemble members does not require many sequential denoising or transport steps. Their main challenge is that CRPS-based objectives are often applied to calibrate the marginal distributions (normalized histogram) of variables, and sometimes of locations and lead times, independently of one another. Marginal calibration reduces the dimensionality of the problem, but it does not by itself guarantee realistic multivariate, spatial, or spectral structure, as the dependence structure across variables and locations (the copula) may remain uncalibrated. 

A second family uses iterative generative samplers. GenCast \cite{price2025probabilistic} uses diffusion-style generation to sample medium-range forecast trajectories, while ATLAS trains stochastic-interpolant and diffusion variants within the same latent transformer framework \cite{kossaifi2026atlas}. These methods define a sequence of transformations from noise or an interpolating process, usually towards the learned conditional distribution. They can produce sharp and coherent samples, but usually result in longer inference times because sampling requires multiple model evaluations.

These families are not strictly exclusive: Direct models can be made more structured, and iterative models can be accelerated. The design of FourCastNet3 \cite{bonev2025fourcastnet3geometricapproach}, for example, attains the spatially correlated, spectrally faithful structure usually associated with iterative samplers while retaining fast direct large-ensemble inference, combining strengths of both families within a single spherical-geometry model. ATLAS also shows that the same backbone can support stochastic interpolants, diffusion, and CRPS-based ensemble training \cite{kossaifi2026atlas}. This clarifies part of the current model landscape: probabilistic skill depends not only on whether a model is ``diffusion-based'' or ``CRPS-trained'', but also on the latent representation, treatment of spatiotemporal spectra, conditioning history, and evaluation of the forecast multivariate state's joint structure. Finally, probabilistic skill also depends on properties inherited from the training data. Models trained on reanalyses can, for instance, exhibit a muted or absent butterfly effect, so that small initial perturbations fail to diverge as they would under the true dynamics \cite{selz2026AIbutterflyeffect}. Such behavior traces back in part to the analyses themselves and to how initial states are estimated, which motivates examining data assimilation directly.

\subsection{From improved data assimilation to end-to-end forecasting}

Most current AI weather models still begin from an analysed atmospheric state. AI can therefore improve forecasting not only by replacing the forecast model, but also by changing how the initial state is estimated \cite{allen2025end,sun_data-to-forecast_2025}. One direction keeps the data-assimilation problem recognizable, but replaces part of the inference methodology. For example, score-based data assimilation \cite{rozet2023score} uses a generative model as a prior over plausible atmospheric states and conditions it on sparse observations, which has been used to produce kilometer-scale surface analyses from weather-station data \cite{manshausen2025generativeda}. 

As this can become expensive for large spatial domains, a promising approach at the global scale is to perform Bayesian data assimilation in a learned latent space, where multivariate physical relationships can help spread observational information across variables and regions \cite{fan2026latentDA}.

These two recent approaches retain the central aim of data assimilation--constructing a physically consistent initial state--but move part of the statistical estimation problem into machine-learned representations.

A more radical direction is to rethink the full observation-to-forecast pipeline. End-to-end observation forecasting systems such as GraphDOP \cite{alexe2024graphdopskilfuldatadrivenmediumrange} and Aardvark Weather \cite{allen2025end} aim to learn directly from heterogeneous observations how to forecast variables of interest, rather than relying on a hand-designed sequence of observation operators, data assimilation, model integration, and post-processing. Instead of first mapping observations into a conventional model state, the model can learn a representation in which observables and forecast variables are directly connected.

The automatic differentiability inherent to deep learning models adds another capability. Traditional tangent-linear and adjoint models have long been central to sensitivity analysis and data assimilation in numerical weather prediction \cite{errico1997adjoint,doyle2014moistadjoint}, but they are expensive to build and maintain for complex physics-based systems. Neural forecasters are differentiable by construction, so backpropagation can provide gradients of forecast quantities with respect to initial conditions or inputs \cite{vonich2024predictabilitylimit2021HW}. Recent studies show that such gradients can recover physically meaningful sensitivity structures for extreme events, with large computational savings relative to traditional adjoint workflows \cite{bano2025aiadjoint}. These gradients do not remove the nonlinear limits imposed by chaotic dynamics, but they facilitate sensitivity experiments, observing-system design, and initial-condition studies.

\subsection{Zooming in: Regional modeling and forecast refinement}

Global AI forecasts are useful, but many hazards depend on scales that they do not resolve directly. A 25-30~km global forecast may capture the synoptic environment for a tropical cyclone, severe convective outbreak, or mountain precipitation event, while still missing the local wind, rainfall, or intensity structure needed for decision-making. AI approaches follow two broad strategies to fill this gap.

The first strategy is to evolve the regional atmosphere itself. AI-based limited-area models learn regional dynamics under lateral boundary forcing, much like traditional regional numerical models but at lower inference cost \cite{adamov2025buildingmachinelearninglimited}. Probabilistic versions such as CRPS-LAM show that regional ensemble forecasts can also be trained to optimize proper scores such as CRPS \cite{larsson2025crpslamregionalensembleweather}. Other approaches avoid explicit lateral boundaries by using global-regional grids, such as stretched-grid models \cite{nipen2025stretchedgrid}  or multiscale global-regional graph frameworks \cite{gao2025oneforecast}. This strategy is appealing because it predicts regional atmospheric dynamics, but boundary treatment, rollout stability, and the cost of forecasting many high-resolution variables are limiting factors.

The second strategy is forecast refinement: complete the coarse global forecast first, then learn a conditional distribution $q_{\theta} $ for a derived local or process-specific quantity of interest, now conditioned on the coarse forecast rather than the initial state alone,
\begin{equation}
y_{t+\Delta t}^{(i)}\sim q_{\theta}\left(\cdot\mid \hat{x}_{t+\Delta t}^{(i)},\ldots,x_{t_{\mathrm{init}}},Z\right),
\qquad i=1,\ldots,N_e.
\label{eq:postprocessing_forecast}
\end{equation}
Here $y$ may be a local hazard variable, station observation, high-resolution wind field, precipitation field, or diagnostic process variable, $x_{t_{\mathrm{init}}} $ denotes knowledge of the initial state, and $Z$ denotes auxiliary information such as topography, land-sea fraction, or satellite data. This formulation includes neural distributional post-processing \cite{bulte2025UQdatadrivenweathermodels,rasplerch2018NNpostprocessing}, hazard-specific post-processing, and weather downscaling. The key point is that the global model need not resolve the hazard explicitly to be useful: it may still predict the large-scale environment from which a specialized model can infer local risk.

\begin{figure}
\centering
\includegraphics[width=0.9\linewidth]{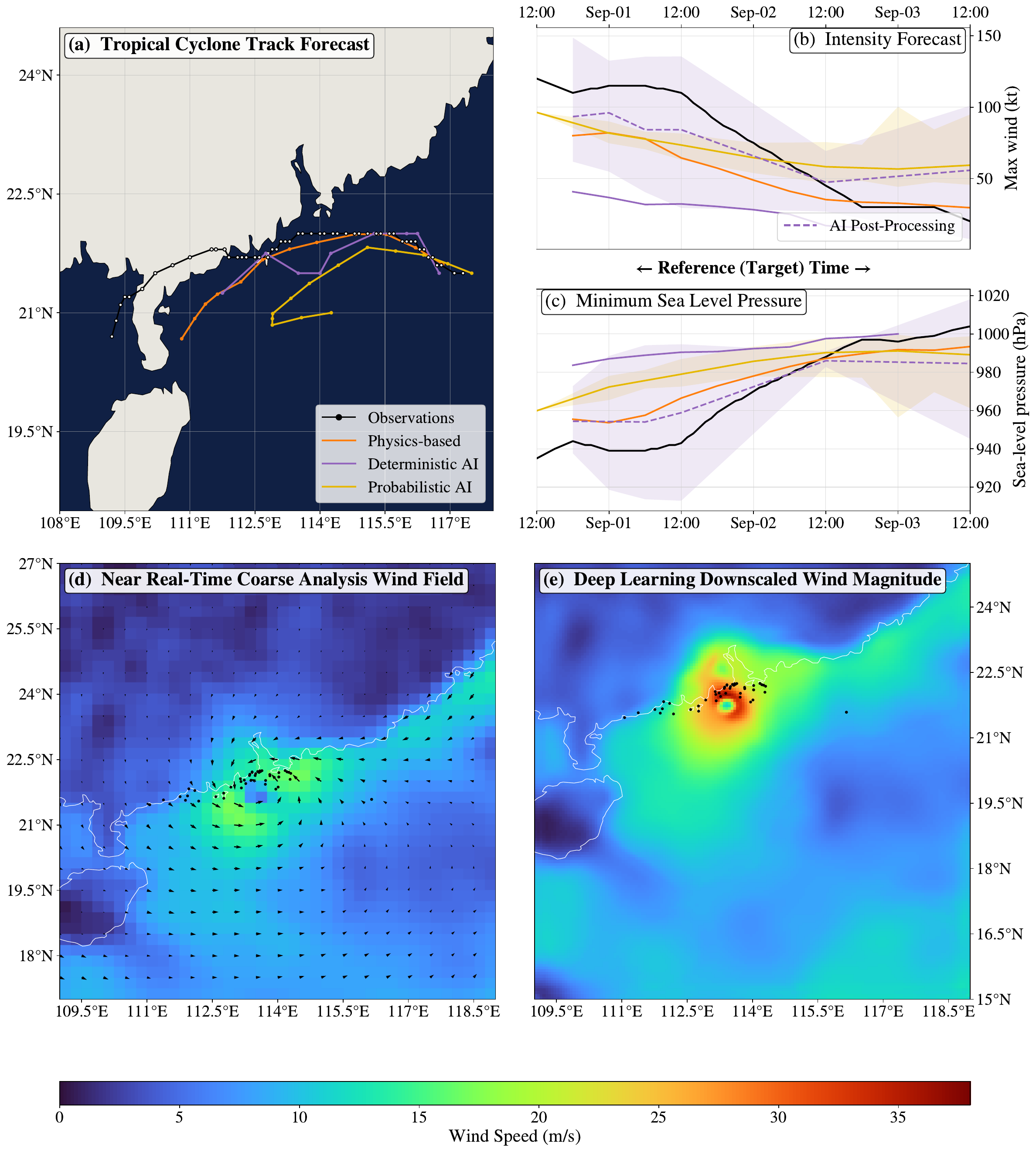}
\caption{\textbf{AI forecast refinement for tropical cyclones, applied to Typhoon Saola (2023).} (a) Track forecasts compared with IBTrACS observations. (b) Maximum wind speed and (c) minimum sea-level pressure forecasts. Black denotes observations; orange the ECMWF IFS ensemble mean; solid purple Pangu; yellow the GenCast ensemble mean; and dashed purple the Pangu post-processing ensemble mean. Shading in (b)--(c) shows the full memberwise range for Pangu and GenCast post-processing. (d) Near-real-time CCMP analysis and (e) deep-learning-downscaled wind magnitude at 06:00 UTC on 1 September 2023. Colors in (d)--(e) show wind speed in m~s$^{-1}$, and black points indicate available in situ observations. Panels (a)--(c) use \cite{gomez2026tcbench}; panels (d)--(e) use \cite{zhang2025NNfusionTC}.}
\label{fig:Saola_Forecast}
\end{figure}

This is already demonstrated for severe weather. Deterministic AI forecasts can contain useful information about convective environments even when convection itself is unresolved \cite{feldmann2024AIconvectiveenv,hua2025Pangufortornadicenv}. Post-processing can then specialize the prediction problem to the hazard, variable, and region of interest, as shown for convective wind gusts \cite{leclerc2025improving}, severe-weather probabilities \cite{hua2026transformerPPofAIWP}, and tropical-cyclone intensity \cite{gomez2025globalforecastingtropicalcyclone}. Tropical cyclones are a useful test case because track, intensity, and local wind structure probe different parts of the forecast system, as illustrated in Fig.~\ref{fig:Saola_Forecast} for Typhoon Saola (2023). Using the TCBench benchmark \cite{gomez2026tcbench}, we compare forecasts initialized at 12:00~UTC against reference observations from the International Best Track Archive for Climate Stewardship (IBTrACS; \cite{IBTrACS}): the ensemble mean from the European Centre for Medium-Range Weather ``TIGGE'' Forecasts \cite{TIGGE}; the deterministic AI forecaster Pangu-Weather \cite{bi2023accurate}; and the probabilistic AI forecaster GenCast \cite{price2025probabilistic}. While tracks are broadly realistic for the first few forecast days (Fig.~\ref{fig:Saola_Forecast}a), AI models struggle to represent intensity directly, consistent with tropical-cyclone intensity biases in ERA5, on which these models are typically trained \cite{bourdin22TCinERA5}. In this setting, a lightweight distributional post-processing model recovers much of the forecasting skill for maximum wind speed (Fig.~\ref{fig:Saola_Forecast}b) and minimum sea-level pressure (Fig.~\ref{fig:Saola_Forecast}c) from the forecast $0.25^{\circ}$ environment \cite{gomez2025globalforecastingtropicalcyclone}, confirming it contains hazard-relevant information. Downscaling based on local observations such as synthetic-aperture-radar winds (Fig.~\ref{fig:Saola_Forecast}d) can complement this strategy: these observations can serve as targets for TransUNet to downscale 25-km Cross-Calibrated Multi-Platform (CCMP) ocean-surface winds for tropical cyclones by factors of 4--8, producing gap-free wind fields and reducing biases in tropical-cyclone maximum wind speed and radius of maximum wind (Fig.~\ref{fig:Saola_Forecast}e) \cite{zhang2025NNfusionTC}. Together, these examples illustrate the role of forecast refinement: the coarse, often global product provides large-scale context, while a specialized model trained on sparse but high-resolution observations reconstructs local hazard structure that the coarse model cannot directly resolve.

\subsection{Foundation models}

Foundation models test a different scaling hypothesis: broad pretraining across variables, vertical levels, datasets, and tasks may produce representations that transfer more readily than single-task forecasting systems. ClimaX \cite{nguyen2023climaxfoundationmodelweather} was an early effort to generalize this idea to weather and climate modeling by training on a combination of reanalysis and climate-model simulations, then adapting the model to tasks such as weather forecasting across lead times, downscaling to reanalysis-data resolution, and emulation of physical climate model projection. Aurora \cite{bodnar2025foundation} and Prithvi WxC \cite{schmude2024prithviwxcfoundationmodel} extend this direction with broader atmospheric datasets and downstream tasks, including atmospheric chemistry and extreme-weather prediction.

A central postulate is that data diversity can improve transfer when the added datasets are physically informative and sufficiently aligned with the target task. Aurora's data-diversity experiments \cite{bodnar2025foundation} provide indications of this: adding datasets generally improves many variables and levels, but adding poorly aligned or insufficiently trained data mixtures can also hurt some targets. Thus foundation models may help integrate reanalyses, forecasts, climate simulations, chemistry, satellite products, and eventually ocean or land data into shared representations, but they do not guarantee tail behavior, physical consistency, and robustness outside the training distribution.

For practical use, foundation-model representations may therefore still require process-specific heads or post-processing. For example, post-processing latent representations learned by foundation models is an emerging way to target variables such as evapotranspiration or precipitation \cite{lehmann2026finetuningweatherfoundationmodel}, where physical and statistical constraints can be imposed more directly. Future development will likely focus less on scale alone than on diverse data fusion, architecture choices, multi-resolution training, and evaluation protocols that reveal when broad pretraining improves generality and when it simply transfers biases across tasks. The forecasting problems above all sit at weather timescales; we now turn to the subseasonal-to-seasonal range, where skill increasingly depends on slowly varying components of the Earth system.

\section{Data-informed subseasonal-to-seasonal forecasting}

Subseasonal-to-seasonal (S2S) forecasting occupies the middle ground between weather and climate prediction. At 2--6-week lead times, forecasts remain sensitive to atmospheric initial conditions, but skill increasingly depends on slower components of the climate system, including sea-surface temperature, the upper ocean, sea ice, the stratosphere, and planetary-scale modes such as the Madden--Julian Oscillation (MJO) and the quasi-biennial oscillation. S2S forecasting therefore lies in a ``predictability desert'': neither atmospheric initial conditions nor slowly varying boundary conditions provide a dominant and reliable source of skill.

What S2S forecasting shares with longer-term climate prediction is that useful forecasts are necessarily probabilistic. Once the deterministic predictability of individual atmospheric states---the first-kind predictability arising from the initial state of the atmosphere \cite{Lorenz75}---is largely exhausted, skill comes instead from shifts in the likelihood of aggregate outcomes: the probability that seasonal temperature or accumulated precipitation exceeds a given threshold. Some predictability persists through slowly varying components such as the ocean and sea ice, whose own initial states remain informative, but even this decays with lead time, and a well-calibrated forecast should then revert to the underlying climate statistics, which may themselves be nonstationary under external forcing such as global warming. S2S predictions are in this sense the shortest forecasts that share the essential character of climate predictions, and they pose a corresponding test: a model skillful at a few days may still fail to reproduce the climate statistics its long rollouts should converge to. This is the first place in our progression from weather to climate where getting the long-run distribution right becomes decisive.

This makes S2S prediction a useful test for AI. Weather-trained autoregressive models can be rolled forward beyond medium range, but the processes that create memory may be implicit, incomplete, or absent when ocean, sea-ice, land, or stratospheric information is simplified. Many ML studies target individual indices or regional variables, including seasonal monsoon precipitation \cite{huang2024skillful}. Here we focus on larger-state-space systems and benchmarks that test whether AI can predict broad atmospheric or Earth-system fields at S2S lead times.

\subsection{Towards objective benchmarking at S2S scales}

The first S2S AI challenge, organized by the World Meteorological Organization in 2021, provided an early community benchmark by asking teams to improve probabilistic 2-m temperature and precipitation forecasts at weeks 3--4 and 5--6 relative to climatology and a recalibrated ECMWF benchmark \cite{vitart2022improvingS2S}. It showed the promise of combining dynamical-model output in a data-driven fashion, but progress remained modest and underlined the need for comparison against strong bias-corrected dynamical baselines. ChaosBench developed an open S2S benchmark \cite{nathaniel2024chaosbenchmultichannelphysicsbasedbenchmark} that includes atmospheric, ocean, land, and ice variables and evaluates models from deterministic and probabilistic viewpoints. It reached a similar conclusion: models developed for weather-scale prediction lose skill rapidly at S2S lead times, often underperforming climatology after about 10 days--the point at which a forecast must earn its skill from the climate distribution rather than the initial state.

The 2025 ECMWF AI Weather Quest moves S2S benchmarking closer to an operational, real-time setting \cite{loegel2025AIWeatherQuest,chantry2025son}. Teams submit weekly probabilistic forecasts for 2-m temperature, accumulated precipitation, and mean sea-level pressure, expressed as probabilities that quintiles of the climatological distribution will be occupied for week-3 and week-4 targets and evaluated with ranked probability skill scores. Post-processing or hybrid AI--dynamical methods have performed strongly: probabilistic bias correction builds on earlier adaptive bias-correction ideas \cite{mouatadid2023adaptive} and has emerged as a strong S2S strategy when applied to both AI and dynamical forecasts \cite{guan2026probabilisticBiasCorrectionS2S}. Among largely data-driven systems, FuXi-S2S scored highest at the time of writing, motivating its use as our first case study.

\subsection{From weather to S2S: The FuXi-S2S approach}

FuXi-S2S illustrates one route from weather to S2S: extend an autoregressive AI weather strategy, then modify it to represent slower predictability sources and forecast uncertainty at 2--6-week lead times \cite{chen2024machine}. The model produces global daily forecasts to 6-week lead times, covering five upper-air variables across 13 pressure levels and 11 surface variables. It is trained from ERA5 daily statistics and evaluated against ECMWF S2S reforecasts using deterministic, probabilistic, MJO, and extreme-event diagnostics.

The key methodological change is not simply a longer rollout. FuXi-S2S adds a perturbation module that injects flow-dependent perturbations into the model's latent space. This differs from conventional ensemble construction, where uncertainty is often represented through perturbations to the initial state. In FuXi-S2S, hidden-state perturbations are learned from historical data and sampled during inference, allowing the model to generate large ensembles efficiently while maintaining forecast spread. This avoids collapsing toward a single trajectory as deterministic predictability decays. The perturbation module is central to its improved probabilistic skill and to its performance relative to ECMWF S2S for variables such as total precipitation and outgoing longwave radiation \cite{chen2024machine}.

FuXi-S2S also improves the representation of low-frequency modes that carry S2S predictability. Its most prominent result is for the MJO, where skillful prediction is extended from about 30 to 36 days. The model better maintains MJO amplitude and phase and captures realistic MJO teleconnections, which helps explain its improved global precipitation skill. This suggests that an autoregressive AI model trained on atmospheric fields can learn part of the organized tropical variability that links initial conditions, slowly varying boundary conditions, and extratropical predictability at S2S lead times.

In its 2022 Pakistan floods case study, FuXi-S2S predicted extreme rainfall anomalies earlier than ECMWF S2S. The authors then used backpropagation of a custom precipitation-anomaly loss to generate saliency maps with respect to sea-surface temperature. These maps highlighted precursor patterns in the tropical central Pacific and western Indian Ocean, consistent with known physical drivers of the event and reassuring from a mechanism perspective.

\subsection{From climate to S2S: Lessons from CliMA-S2S}

Instead of extending an AI weather model toward longer lead times, CliMA starts from a hybrid climate-modeling philosophy and uses S2S prediction as an objective development test. CliMA retains a physics-based dynamical core for the resolved climate system while adding data-driven components for selected parts of the modeling pipeline \cite{yatunin2026CliMA_dycore}: data-driven closure models for source and sink terms within parameterizations, AI flux corrections for aggregate subgrid fluxes, offline post-processing and bias correction of model output against observations, and emulators mapping uncertain parameters to model diagnostics for uncertainty quantification.

\begin{figure}
    \centering
    \includegraphics[width=\linewidth]{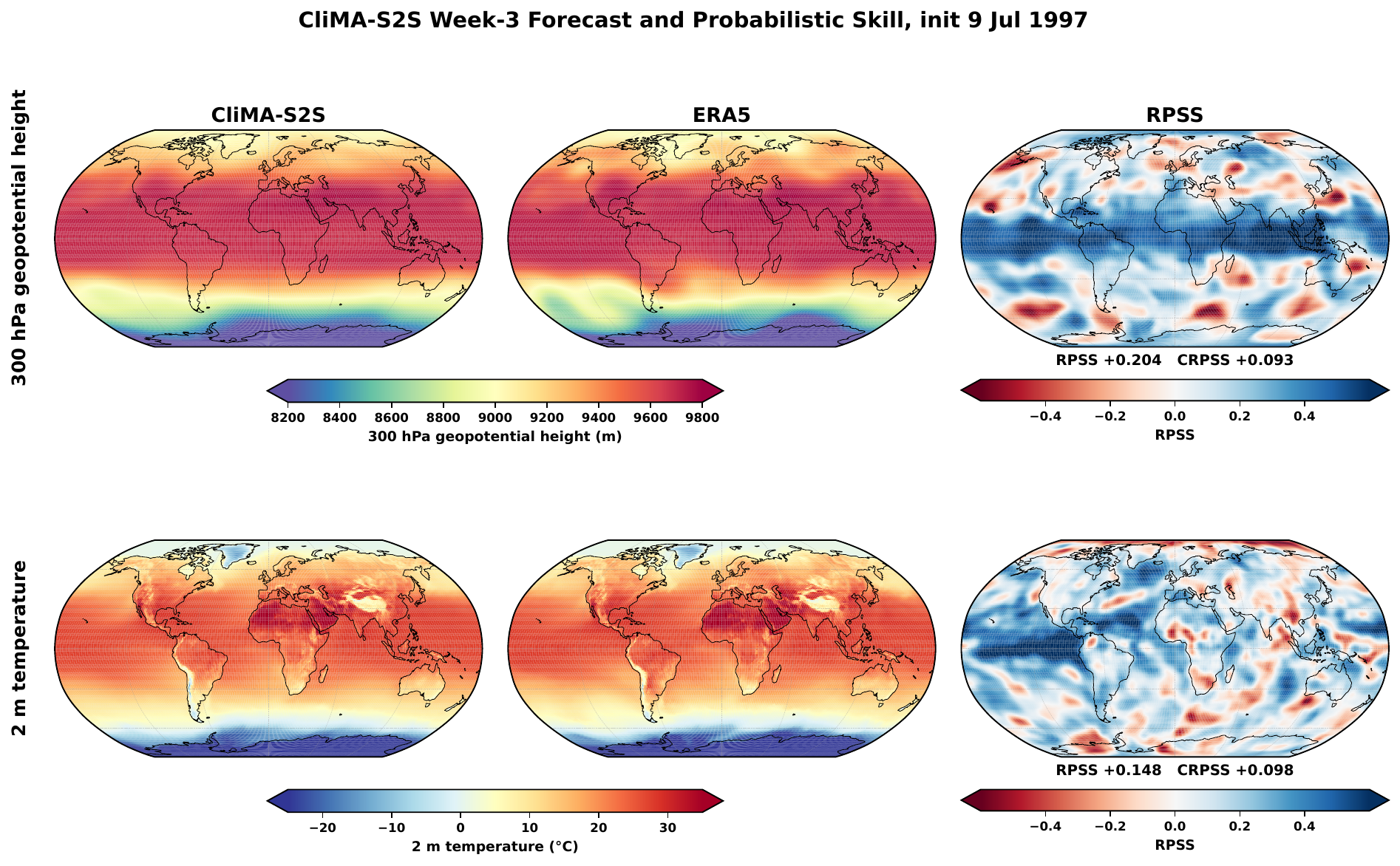}
    \caption{\textbf{CliMA-S2S week-3 forecasts and probabilistic skill.} Seven-day-mean 300-hPa geopotential height (top) and 2-m temperature (bottom) for lead days 15–21, initialized 9 July 1997. Left panels show the bias-corrected CliMA-S2S forecasts, middle panels show the corresponding ERA5 verification fields, and right panels show ranked probability skill score (RPSS) for climatological quintiles relative to ERA5 climatology. Because CliMA-S2S is run here as a single deterministic forecast, predictive uncertainty is estimated through lightweight leave-one-out statistical calibration rather than from an ensemble of perturbed initial conditions. Positive RPSS indicates improvement over climatology. Global RPSS and continuous ranked probability skill score (CRPSS) are indicated below the right panels.}
    \label{fig:clima-s2s}
\end{figure}

This makes S2S evaluation useful in a different way from FuXi-S2S: it provides a more direct test of short-time climate dynamics than time-mean climate statistics alone. To achieve CliMA's flagship AMIP setup, the atmosphere model was developed and evaluated using seasonal-to-subseasonal runs of roughly 40 days or less, including through participation in the AI Weather Quest. The S2S and climate-development objectives overlap for several core variables: mean sea-level pressure, 2-m temperature, surface precipitation, and geopotential height. A representative week-3 forecast from  ``CliMA-S2S'', the CliMA coupled land–atmosphere model with postprocessing, is shown in Fig.~\ref{fig:clima-s2s}  following a 9 July 1997 initialization. Starting from a single deterministic coupled-model trajectory, lightweight distributional post-processing corrects systematic forecast biases and estimates predictive uncertainty from ERA5-based calibration statistics, rather than from an ensemble of perturbed initial conditions. The resulting probabilistic forecasts capture the large-scale structure seen in ERA5 for 300-hPa geopotential height and 2-m temperature and yield positive probabilistic skill relative to climatology. Beyond these forecast variables, climate-model development uses process-specific diagnostics evaluated against observational products, including liquid water path against MAC-LWP  over ocean \cite{elsaesser2017MACLWP}, top-of-atmosphere cloud radiative effects against CERES \cite{loeb2018CERES}, and three-dimensional cloud fraction against CloudSat/CALIPSO \cite{bertrand2024CloudSatCALIPSO}.

The Weather Quest setting exposed concrete axes of improvement for climate-model development. First, stratospheric circulation biases motivated changes to the model top and sponge-layer treatment. Second, sea-ice initialization mattered: initializing only with a constant freezing condition, rather than a two-dimensional sea-ice state, produced polar 2-m temperature and mean-sea-level-pressure biases that persisted beyond the first forecast week. Third, gravity-wave treatment produced biases visible within a week in surface pressure and in stratospheric and upper-tropospheric momentum, with effects that also persisted beyond the first week. Many of the resulting choices were retained in later coupled CliMA configurations because they improved model behavior.

This highlights why S2S can be valuable for developing climate models. Climate-model calibration is often an inverse problem: parameters are adjusted so that long-run conditional statistics, such as cloud or radiation metrics, match observations despite internal variability. While S2S weekly forecasts are too noisy and initial-condition dependent for fully learning some fields, such as precipitation, they can still be informative for smoother fields such as mean sea-level pressure. This tension between short-range forecast signals and the long-run statistics one ultimately wants to constrain recurs throughout climate modeling, and we return to it when discussing how models should be selected and evaluated.

\section{AI in Climate Modeling}

\subsection{What must an Earth system model represent?}

The central question in moving from weather to climate modeling is not only whether an AI model can sustain long rollouts, but whether it can produce reliable responses to unseen, time-dependent forcings. For climate projection, different modeling strategies must ultimately address three coupled questions. First, how will climate---including clouds, circulation and extreme events---change as greenhouse gases such as CO$_2$ and CH$_4$ rise? Second, how will aerosol-emission changes affect climate through direct radiative effects and indirect effects on cloud formation? Third, how will carbon-cycle feedbacks partition emitted CO$_2$ among the atmosphere, land biosphere, and ocean? The expected output is the temporal evolution of spatially resolved physical fields, from which equilibrium climate sensitivity (ECS), transient climate response (TCR), and transient climate response to cumulative carbon emissions (TCRE) can be derived.

Climate attribution also requires counterfactual historical scenarios, such as the climate had greenhouse-gas concentrations not increased. Such counterfactuals are embodied in the Coupled Model Intercomparison Project scenario sets, including ScenarioMIP for CMIP6 \cite{ONeill2016ScenarioMIP,riahi2017sharedSSP} and CMIP7 \cite{van_vuuren2026scenarioMIP_CMIP7}. How to produce them causally remains an open question. Direct causal-emulation strategies are emerging, exemplified by PICABU \cite{hickman2025causalclimateemulationbayesian}, which constrains each grid cell to depend on a single latent climate mode and learns sparse lagged links among these modes, enabling targeted intervention experiments and counterfactuals. More generally, relevant forcing agents must be independently manipulable; representing them explicitly through physically-informed pathways is one way to achieve this. Let $\varphi^{(i)}$ denote the prognostic state of realization $i$, where realizations may differ in their forcing trajectories or initial conditions:

\begin{equation}
\frac{\partial \varphi^{(i)}}{\partial t} ={\cal F}^{(i)}\left(\varphi^{(i)},t\right),\qquad\varphi=\left[\underbrace{T,q,\boldsymbol{u},\ldots}_{\varphi_{\mathrm{atm}}},\varphi_{\mathrm{ocean}},\varphi_{\mathrm{land}},\varphi_{\mathrm{cryo}},\ldots\right]^{\mathsf{T}} .
\label{eq:climate_state_tendency}
\end{equation}

Here, ${\cal F}^{(i)}$ is the state-tendency operator; $T$, $q$, and $\boldsymbol{u}$ denote atmospheric temperature, specific humidity, and wind; and $\varphi_{\mathrm{atm}}$, $\varphi_{\mathrm{ocean}}$, $\varphi_{\mathrm{land}}$, and $\varphi_{\mathrm{cryo}}$ collect the prognostic variables of the atmosphere, ocean, land, and cryosphere, respectively. 

\begin{figure}
\centering
\includegraphics[width=\linewidth]{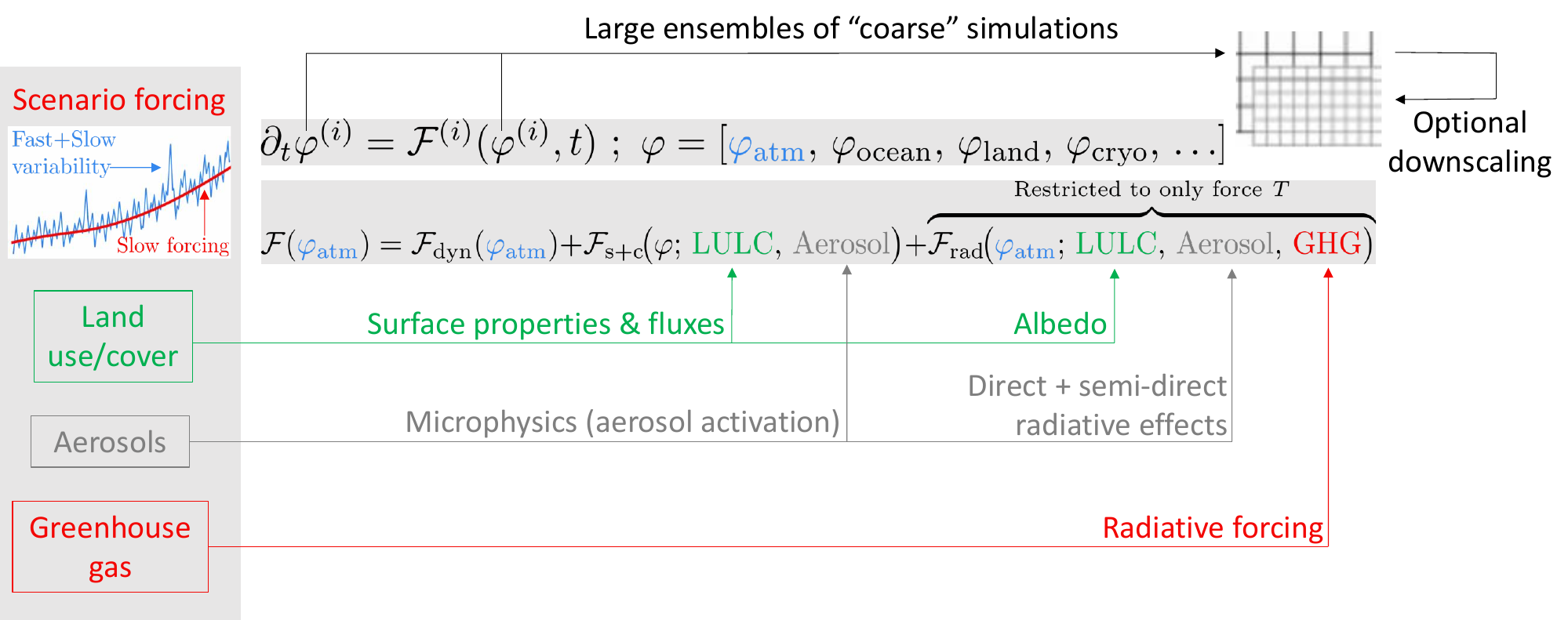}
\caption{\textbf{Example formulation for including external forcing agents through restricted pathways, illustrated here for atmospheric tendencies.}  Land use and land cover (LULC) enter the subgrid/coupling term through surface properties and fluxes and the radiative term through albedo; aerosols enter the subgrid/coupling term through microphysics and the radiative term through direct and semi-direct effects; and greenhouse gases (GHGs) enter only through radiative forcing. Here, the direct atmospheric radiative tendency acts only on temperature, while other variables respond through subsequent dynamics, subgrid processes, and coupling. These pathways may become intertwined over multi-hour model steps, while persistent forcing modifies the evolving climate and its internal variability. Multiple realizations, denoted with superscript (i), for a given external forcing can generate large ensembles of coarse simulations, with optional downscaling; emissions-driven Earth system models must additionally represent the chemistry and biogeochemistry that determine GHG and aerosol concentrations. }
\label{fig:minimum_climate_model_formulation}
\end{figure}

Anthropogenic and natural forcing agents, including greenhouse gases and aerosols, affect climate primarily through radiative heating, which contributes directly to temperature tendencies in the atmosphere and the surface layers of land and ocean. A minimal causal requirement is therefore that relevant agents enter as explicit, independently manipulable inputs whose effects remain identifiable when varied separately. Fig.~\ref{fig:minimum_climate_model_formulation} illustrates one way to make these pathways explicit, while autoregressive AI models with time steps of six hours or more may instead mix radiative heating with other processes within a learned update. In concentration-driven configurations, greenhouse-gas and aerosol concentrations can be prescribed; emissions-driven Earth system models must instead predict them by representing atmospheric chemistry and land and ocean biogeochemistry.

Earth system processes span overlapping timescales. Long-lived greenhouse gases exert small but persistent radiative perturbations that slowly modify weather and internal climate variability, whereas aerosol--cloud interactions and cloud feedbacks can evolve on weather timescales. Learned updates may capture such interactions \cite{mahesh2026fastRadiativeFeedbacks}, but slow ocean and atmospheric-composition feedbacks extend into climate states without recent historical analogues. Land- and water-use changes similarly act through slowly evolving boundary conditions and changes in surface albedo, roughness, soil moisture, and energy and water fluxes.

Two broad modeling strategies follow: Autoregressive pure AI models learn most of a component (e.g. the full atmosphere or ocean) or coupled-system update operator. Such pure AI models can be used to efficiently emulate reference physical model trajectories. Climate emulators vary in complexity \cite{balaji2017CPMIP}; we here focus on emulators that represent spatially resolved climate fields, simulate weather-scale variability, and could support future AI Earth system models \cite{henn2026AIMIPPhase1}. Hybrid AI--physics models prescribe physical equations for processes such as resolved fluid dynamics and radiative transfer while learning unresolved or uncertain processes from data, particularly subgrid-scale parameterizations \cite{Schneider17c,eyring2024ai}.

Both strategies are limited by the scarcity of observed forced responses. The historical record contains one realization of the slow forced response, with forcing agents covarying, so pure AI models trained on observations or reanalyses may learn correlations that fail in counterfactual climates. Controlled simulations can broaden emulator training and enable direct OOD tests against held-out reference-model simulations, but such tests establish fidelity to the reference, not to Earth. Physically based models likewise cannot be verified against unobserved climates, and their responses remain uncertain, as illustrated by the CMIP6 spread in ECS \cite{meehl2020ECS_CMIP6}.

Most current AI emulators represent fewer processes than mature Earth system models. Emissions-driven simulations require interactive carbon-cycle and atmospheric-composition processes, including land and ocean biogeochemistry, atmospheric chemistry, and aerosol processes. Replacing components within mature models offers hybrid AI-physics models a near-term route to participation in CMIP7, while coupled emulators need to incorporate additional components. Development of AI emulators should therefore complement continued development of process-based climate and Earth system models as both strategies mutually benefit each other. Beyond these strategies, AI can be combined with observations to refine estimates of variables such as effective radiative forcing \cite{vanloon2025obs_based_estimates_eff_rad_forcing}, and to design controlling-factor analyses that hold across climates \cite{nowack2025emergent_constraints}.

\subsection{AI climate-model benchmarking and AIMIP Phase 1}

The development of AI climate models requires evaluation beyond short-range forecast accuracy. Indeed, most autoregressive AI climate models are optimized using weather-prediction losses over lead times of a few days, whereas climate performance is defined based on statistics from multi-year or multidecadal simulations \cite{ullrich2024recommendations_for_eval}. Lower short-range forecast error does therefore not guarantee smaller climate bias or a more accurate forced response, and long-rollout diagnostics and climate statistics must therefore be included in model selection and evaluation.

ClimateBench provided an early benchmark for emulating global and spatially resolved forced climate responses from prescribed emissions and concentrations \cite{watson2021ClimateBench}. AIMIP extends this effort to temporally-resolved atmospheric models that explicitly generate weather variability. Its first phase establishes a common AMIP-style framework, output format, and training constraints for evaluating AI weather and climate models as atmospheric climate models \cite{henn2026AIMIPPhase1}.

AIMIP Phase 1 models simulate the atmosphere over 1979--2024 under prescribed historical sea-surface temperatures. The evaluation considers mean-climate biases, historical trends, atmospheric responses to El Ni\~no-related sea-surface-temperature anomalies, temporal variability, and OOD generalization. The common outputs are provided in CMIP-compatible formats, allowing the wider climate-modeling community to apply established process-oriented and regional diagnostics.

The first results show that several AI models reproduce major features of the historical atmosphere and its response to prescribed El Ni\~no variability. However, some models underestimate historical warming trends, and model responses diverge in OOD experiments, including spatially uniform sea-surface-temperature perturbations \cite{henn2026AIMIPPhase1}. Historical AMIP performance is therefore a necessary but insufficient test of climate-model credibility. Models must also respond appropriately to boundary conditions and external forcings outside the training distribution.

Therefore, AIMIP, ClimateBench, and broader Earth system AI benchmarking efforts within the AICMIP umbrella \cite{luo2026AIatm_oce}, complement but do not replace the experimental hierarchy being developed for CMIP7. Future evaluation should build upon the measures of mean climate, internal variability, trends, extremes, and transient responses to prescribed forcing that they established. In particular, assessments of top-of-atmosphere and atmospheric energy budgets, atmospheric and oceanic mass and water budgets, and emergent diagnostics such as evaporation minus precipitation would be valuable, especially in the context of models coupling multiple Earth system components together. Depending on model scope, forced-response tests should include the sign and magnitude of ECS, TCR, and TCRE.

\subsection{Hybrid physics-AI Earth system modeling}

\subsubsection{From hybridizing existing models to constraining AI models}

Hybrid atmospheric models typically retain a dynamical core while augmenting selected processes with data-driven components. For instance, the Neural General Circulation Model (GCM) combines a fully differentiable atmospheric dynamical core with learned parameterizations \cite{kochkov2024neural}, allowing direct training against satellite observations for variables such as precipitation, whose diurnal statistics remain challenging to model in conventional GCMs \cite{yuval2026neural}. The ML-enhanced (MLe) version of ICON-A confirmed the advantages of data-driven equation discovery for parameterization \cite{zanna2020eqdisc}, provided that the learned parameterization is jointly recalibrated with existing ones, improving agreement with observations of the target process after recalibration \cite{grundner2025ICONMLe}. Data-driven ocean closures for hybrid modeling now span the full complexity spectrum, including physics-informed and scale-aware models, facilitating the broader goal of coupled atmosphere--ocean--sea-ice hybrid simulations \cite{zanna2025frameworkhybridphysicsaicoupled}. Related approaches are surveyed in the HybridESM living review \cite{beucler2025HybridESMLivingReview}; we henceforth focus on the Climate Modeling Alliance (CliMA) as it provides a comprehensive example of hybrid climate modeling.

\subsubsection{The CliMA approach}

CliMA combines process-based models that respect physical conservation laws with data-driven components to model the entire Earth system, including atmosphere, oceans, and land. Physical laws and their deductive power provide a scaffolding for out-of-distribution generalizability \cite{Christopoulos_2024}, particularly where training data are unavailable to constrain climate projections. This physical framework is augmented with data-driven models to enhance expressiveness, flexibility, and accuracy where sufficient data exist. All model components are designed to be efficient on modern GPU architectures, which enables the large, distributed ensembles required for calibration and uncertainty quantification. This computational strategy allows the model to be trained against observed climate statistics, a method distinct from learning from trajectories of individual weather states, as done in AI weather forecasting, and appropriate for the climate prediction problem \cite{Schneider17c}.

Training the model with climate statistics is primarily accomplished using derivative-free ensemble Kalman methods \cite{Iglesias13a,Kovachki19a,Huang22a}. These methods approximate stochastic gradient descent through ensembles of simulations and are well-suited for chaotic systems because the ensemble smooths noise-induced extrema in the loss function---extrema that arise from the climate system's internal variability when the loss to be minimized contains finite-time statistical averages of observables such as cloud cover. Because they require neither adjoints nor differentiability of the forward model, they can be applied to legacy components and to non-differentiable parameterizations, and they reach comparable or lower loss than alternative derivative-free optimizers at a fraction of the cost \cite{Gjini26a}. This framework allows learning about small-scale processes, such as turbulent air mass exchange between clouds and their environment, in an inverse problem setting \cite{Lopez-Gomez22a,Dunbar21a,Christopoulos_2024}. In this setting, indirect data such as satellite measurements of cloud cover and liquid water path can be used to constrain processes that are not directly observable.

Calibration alone yields point estimates, but projections require uncertainty quantification of the learned parameters and propagation of these uncertainties into projected climate statistics. CliMA therefore embeds uncertainty quantification in the calibrate–emulate–sample framework \cite{Cleary21a}: the ensembles generated during Kalman inversion are reused to train an emulator of the model–data misfit, which is then sampled by Markov chain Monte Carlo to approximate the Bayesian posterior over parameters. This reduces the number of model evaluations needed for posterior sampling by orders of magnitude relative to direct MCMC, making Bayesian uncertainty quantification tractable for global models \cite{Dunbar21a,Howland22a}. The resulting posterior ensembles can be propagated forward to obtain distributions of climate statistics of interest, including climate sensitivity. The calibration and uncertainty quantification tools are developed as open-source, model-agnostic packages \cite{Dunbar22b}.

This hybrid methodology is exemplified in the parameterization of turbulence, clouds, and convection, a dominant source of uncertainty in climate projections. Cloud systems respond to temperature and circulation changes, as well as to direct changes in greenhouse gas concentrations that modulate the thermal infrared radiation budget of clouds. In the observational record, the thermal and direct radiative effects are confounded, as temperatures and greenhouse gas concentrations have increased simultaneously. A model must be able to predict these effects separately to be credible in scenarios where they might decouple, such as in geoengineering applications. A purely data-driven model trained on historical data could fail this test. CliMA's approach uses a physical scaffolding that explicitly models radiative transfer and enforces conservation of mass and energy. Within this framework, AI-based models learn representations for smaller-scale processes, such as the turbulent exchanges between clouds and their environment, which are critical controls on the cloud response to warming but are difficult to model from first principles \cite{Christopoulos_2024}. These data-driven small-scale models are often pre-trained offline with computationally generated data, such as libraries of high-resolution large-eddy simulations \cite{Shen22j,Chammas26a}, before being fine-tuned with Earth observations in online, global simulations.

A different application of the hybrid strategy is used in aspects of land surface modeling. Processes governing snow thickness and albedo are physically complex. However, extensive observational data exist that link snow properties to environmental variables such as temperature, humidity, and precipitation. It is plausible that the variations of these properties and variables in the present climate—across latitudes and altitudes—span the conditions that will occur in a warmer world. This makes data-driven models for snow processes viable. CliMA employs a data-driven snow model, but embeds it within physical guardrails that ensure conservation of water mass and energy \cite{Charbonneau25a,Charbonneau26a}. The same principle, where data-driven components are constrained by fundamental physics, is applied to other land surface processes, striking a balance between empirical accuracy and physical consistency.

The hybrid design extends across components. The atmospheric dynamical core conserves air mass, water mass, and moist energy even with subgrid-scale parameterizations active, bounding the closures it hosts \cite{yatunin2026CliMA_dycore}; ClimaLand couples water, energy, and carbon fluxes through soil, snow, and canopy \cite{Deck26a}; Oceananigans resolves mesoscale eddies at climate timescales \cite{Silvestri25a}; and ClimaCoupler exchanges surface states and fluxes across grids and timesteps. Trainable closures enter wherever unresolved processes dominate uncertainty, while conservation laws constrain the causal structure, facilitating generalization under altered forcing and for counterfactual scenarios.

\subsection{Autoregressive Earth system emulation}

\subsubsection{AI atmospheric model emulation: Lessons from ACE}

A complementary strategy is to learn most of an Earth system component directly from data. Autoregressive AI weather models cannot necessarily be extended directly to climate timescales. Many models drift from realistic climates or become unstable after several months \cite{Kar2024}. Models trained on reanalysis aim to reproduce the observed climate, its trends, and its variability over the available record, but cannot be expected to generalize automatically to future climate states. Models trained on physically based climate simulations instead aim to emulate a reference model and can be trained across several prescribed climates. At comparable resolution, ACE2 reports approximately $130\times$ higher inference throughput than its C96 SHiELD reference model and approximately $700\times$ lower energy use per simulated year, estimated from runtime and rated hardware power \cite{watt2025ace2}.

For example, DLESyM is trained primarily on reanalysis, predicts the atmosphere autoregressively, and includes a simple interactive ocean representation \cite{cresswellclay2025DLESyM}. ACE \cite{wattmeyer2023ACE1,watt2025ace2} and CAMulator \cite{chapman2025camulator} have been configured to emulate the atmospheric components of reference climate models. The ACE program has developed a sequence of progressively more demanding atmospheric emulation experiments and is therefore discussed in detail below.

The goal of ACE is to emulate a reference global atmospheric model across a range of climates while reproducing its weather variability and climate statistics. The longer-term goal is to extend this approach to the coupled atmosphere--ocean--land--sea-ice system and eventually to other Earth system processes such as biogeochemical cycles. The reference training data must span the forcing combinations over which the emulator is expected to operate.

ACE uses NVIDIA's open-source Spherical Fourier Neural Operator architecture \cite{bonev2023SFNO}, typically at approximately $1^\circ$ horizontal resolution with a 6-hour time step. ACE modifies the standard SFNO architecture to support climate applications and budget calculations. It predicts the surface-flux components needed to calculate atmospheric heat and moisture budgets. It also predicts the atmospheric state in eight terrain-following layers that can be summed into column-integrated heat and moisture without additional vertical interpolation.

ACE has been trained to emulate the GFDL/NOAA's SHiELD model \cite{harris2020GFDL_SHiELD}, a variant of the FV3GFS global weather model, and the atmospheric component of the US Department of Energy's E3SM \cite{golaz2019DOE_E3SM}. The first task used a repeating seasonal cycle with prescribed sea-surface temperatures and insolation \cite{wattmeyer2023ACE1,duncan2024ACE2precip}. In these configurations, ACE reproduced the climates of both reference models. Training required approximately 2.5 days on four NVIDIA A100 GPUs, and inference produced approximately 800 simulated years per wall-clock day on one A100 GPU.

\begin{figure}
\centering
\includegraphics[width=\linewidth]{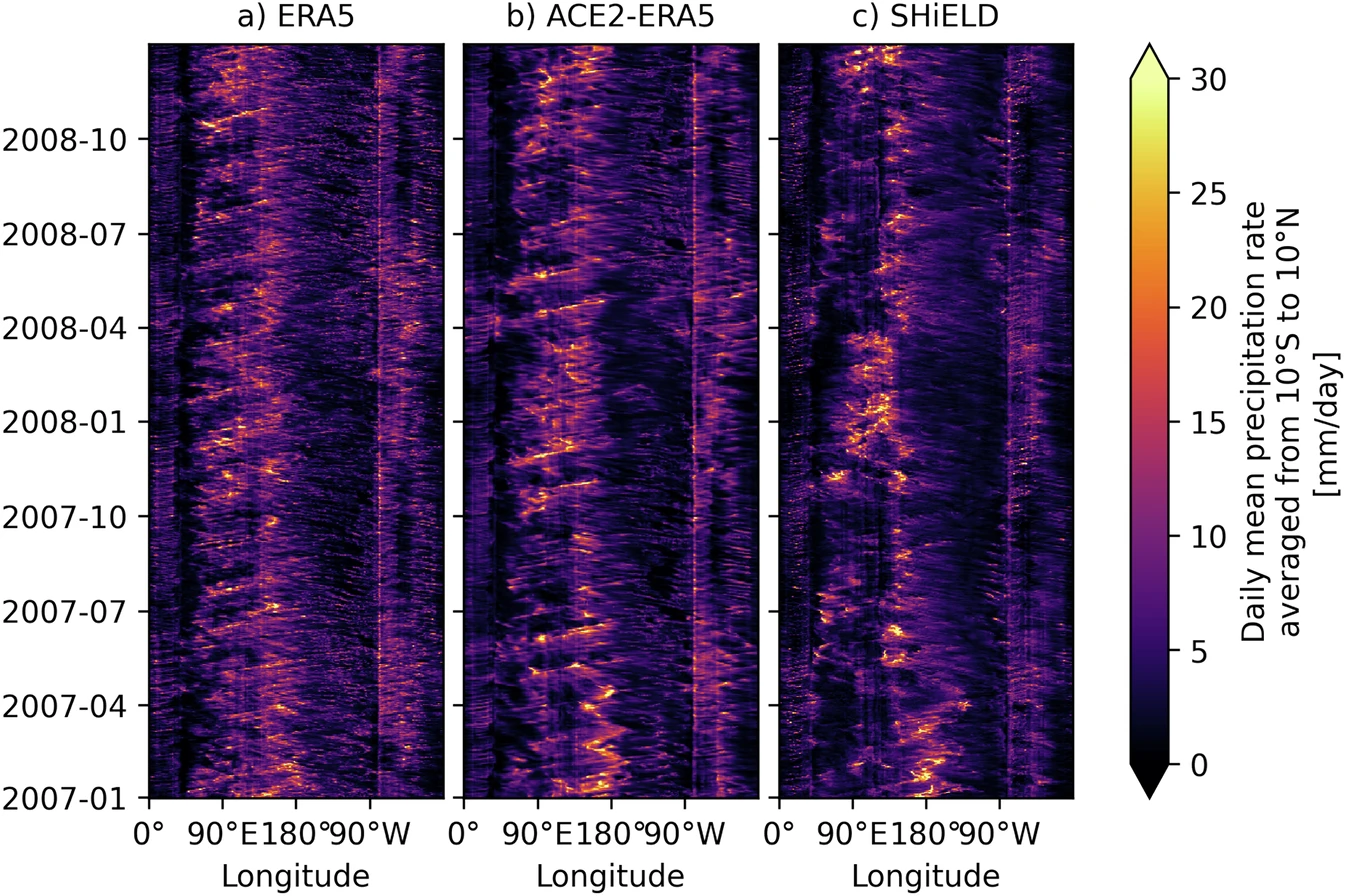}
\caption{\textbf{Tropical precipitation variability in ACE2.}
Two-year Hovm\"oller diagrams of daily precipitation averaged from $10^\circ$S to $10^\circ$N show that a long ACE2 rollout trained on ERA5 reproduces statistical characteristics of ERA5 tropical precipitation, including eastward-propagating variability associated with the Madden--Julian Oscillation. Adapted from Fig.~6 of \cite{watt2025ace2}.}
\label{fig:ACE2}
\end{figure}

The second task was AMIP-style training and testing over the historical period from 1940 to 2020. Century-scale rollouts motivated hard conservation of global dry-air mass and closure of the global atmospheric moisture budget. The resulting ACE2 model reproduces the time-mean climate, global trends, and SST-forced ENSO variability of ERA5 or a reference model \cite{watt2025ace2}. When trained on ERA5, ACE2 also reproduces statistical characteristics of tropical precipitation variability, including the Madden--Julian Oscillation (Fig.~\ref{fig:ACE2}), and the global spatial distribution of tropical cyclones.

\begin{figure}
\centering
\includegraphics[width=\linewidth]{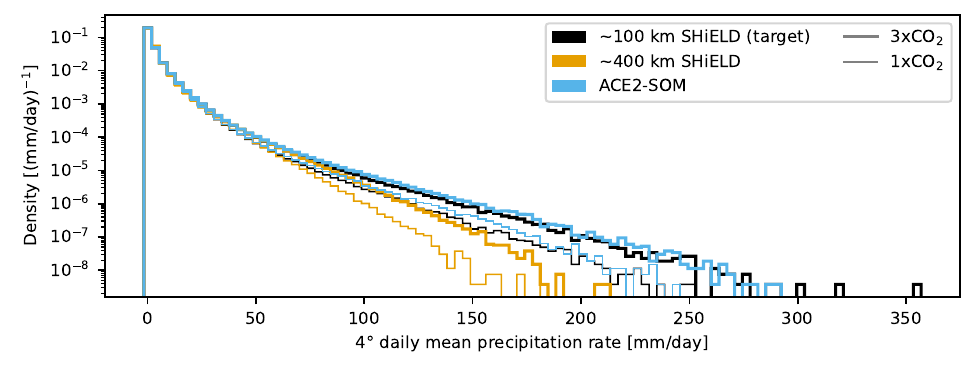}
\caption{\textbf{Extreme precipitation in ACE2-SOM.}
Daily precipitation distributions from 50-year samples of reference slab-ocean-coupled $\sim$100~km and $\sim$400~km SHiELD simulations, and ACE2-SOM for present-day and unseen $3\times$CO$_2$ climates, regridded to a common 4$^{\circ}$ Gaussian grid. ACE2-SOM reproduces the target $\sim$100~km model's extreme-precipitation frequency and its sensitivity to CO$_2$-induced warming. Adapted from Fig.~6 of \cite{clark2024ACE2-SOM}.}
\label{fig:ACE_SOM_1}
\end{figure}

The third task coupled ACE2 to a slab-ocean model and trained the coupled emulator on analogous SHiELD simulations in equilibrium climates with present-day, doubled, and quadrupled CO$_2$ concentrations \cite{clark2024ACE2-SOM}. ACE2-SOM reproduces an unseen intermediate $3\times$CO$_2$ simulation, providing an example of interpolation between climates used for training. This includes changes in means and extremes. In particular, ACE2-SOM reproduces the extreme-precipitation tail and its sensitivity to $3\times$CO$_2$ warming more accurately than a coarser $\sim$400~km version of SHiELD (Fig.~\ref{fig:ACE_SOM_1}).

\begin{figure}
\centering
\includegraphics[width=\linewidth]{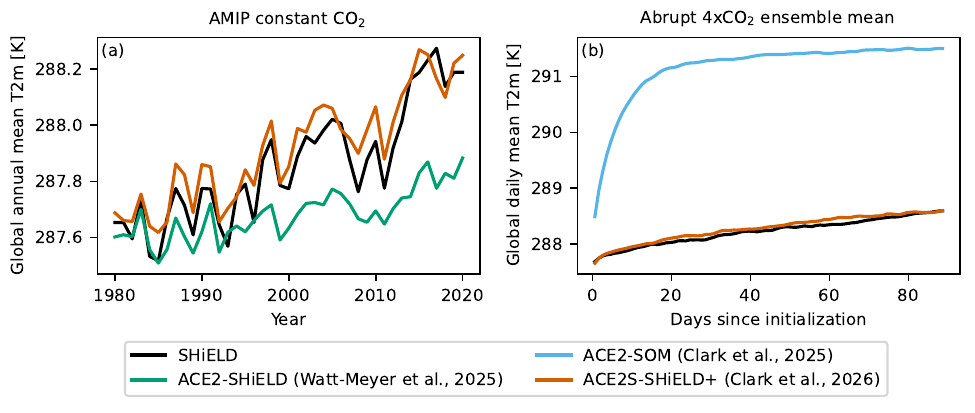}
\caption{\textbf{AMIP constant CO$_2$ and slab-ocean abrupt 4xCO$_2$ experiments with ACE.}
Evolution of global annual mean 2-m temperature in simulations forced by observed sea-surface temperature and sea ice, but with CO$_2$ held constant at the 1979 level with the reference dynamical SHiELD model, ACE2-SHiELD, and ACE2S-SHiELD+ (a); and evolution of global daily and ensemble mean 2-m temperature in a 36-member ensemble of 90-day slab-ocean-coupled abrupt 4xCO$_2$ simulations with SHiELD, ACE2-SOM, and ACE2S-SHiELD+ (b).  ACE2-SHiELD was trained solely on output from historical AMIP simulations; ACE2-SOM was trained solely on output from slab-ocean-coupled equilibrium-climate simulations with 1x, 2x, and 4x the present-day CO$_2$; and ACE2S-SHiELD+ was trained on a mixture of output from AMIP, equilibrium-climate, and ramped-SST, random-CO$_2$ simulations. Adapted from \cite{clark2026ACE2S_SHIELDplus}.}
\label{fig:ACE2S-SHiELD+}
\end{figure}

OOD tests also exposed limitations in models trained solely on data from AMIP or equilibrium-climate simulations, particularly in disentangling the effects of SST and CO$_2$. ACE2-SHiELD forced by observed SST and sea ice, but with CO$_2$ held constant at the 1979 level, exhibited an unrealistically muted trend in global mean 2-m temperature (Fig.~\ref{fig:ACE2S-SHiELD+}a).  In an abrupt 4xCO$_2$ experiment, the atmosphere of ACE2-SOM adjusted to a 4xCO$_2$ equilibrium climate within roughly 90~days, breaking global energy conservation and inconsistent with the slower adjustment of its physics-based ocean (Fig.~\ref{fig:ACE2S-SHiELD+}b). ACE2-SOM's surface-radiative-flux response to changing CO$_2$ at fixed atmospheric state also differed from a radiation double-call in the reference model. These experiments show that interpolation within AMIP or across equilibrium climates does not ensure the correct response when sea-surface temperature and CO$_2$ vary independently.

ACE2S-SHiELD+ addresses this problem by supplementing AMIP and equilibrium-climate simulations with ramped-SST, random-CO$_2$ simulations, which sample many diverse combinations of SST, atmospheric state, and CO$_2$ \cite{clark2026ACE2S_SHIELDplus}. It is evaluated on forcing combinations held out from training, including AMIP constant CO$_2$ (Fig.~\ref{fig:ACE2S-SHiELD+}a), AMIP $+4$~K, held-out random-CO$_2$ trajectories, and slab-ocean abrupt-$4\times$CO$_2$ (Fig.~\ref{fig:ACE2S-SHiELD+}b) experiments. ACE2S-SHiELD+ retains comparable performance to the earlier ACE2-SHiELD and ACE2-SOM models in their strongest historical AMIP and equilibrium-climate tests, improves the separation of SST-mediated and direct CO$_2$ responses, and is trained on roughly 25\% fewer samples than either alone. Exposing ACE to aerosols or other greenhouse gases as independent forcings would bring it one step closer to conventional climate models.

\begin{figure}
\centering
\includegraphics[width=\linewidth]{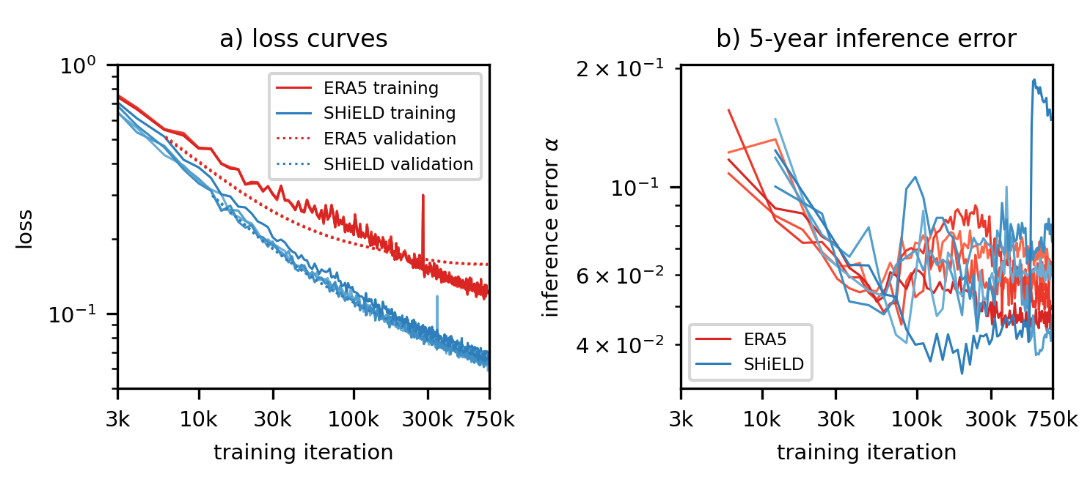}
\caption{\textbf{Short-range loss and long-range climate error in ACE2.}
Training and validation losses for ACE2-ERA5 and ACE2-SHiELD are compared with an inference error computed from eight five-year simulations after each epoch. Short-range training loss and long-range climate error do not necessarily select the same model checkpoint. Adapted from Fig.~S8 of \cite{watt2025ace2}.}
\label{fig:ACE2_ERA5_SHIELD}
\end{figure}

A common difficulty for autoregressive atmospheric emulators, related to the fast vs. slow forcings, is that their training loss measures weather prediction over lead times of at most a few days, whereas the relevant evaluation metrics are statistics from multi-year or multidecadal simulations \cite{ullrich2024recommendations_for_eval}. Lower short-range forecast error does not guarantee lower climate bias (e.g. Fig.~\ref{fig:ACE2_ERA5_SHIELD}). One practical approach is to evaluate long climate rollouts during training and select checkpoints using climate statistics rather than short-range loss alone.

\subsubsection{AI ocean climate modeling}

Amongst all components of the Earth system, ocean representation is particularly important for climate feedbacks because ocean heat uptake, circulation, and carbon storage determine major components of the transient and equilibrium climate response. A surface mixed layer may be sufficient for some idealized atmospheric feedback experiments, but coupled climate projections require representations of ocean circulation, deep-ocean heat uptake, and carbon storage. Learned ocean models have been developed for short-range forecasting, monthly-to-decadal prediction, and multicentury climate emulation.

\paragraph{Developing learned ocean components}

By varying the prognostic variables, depth, horizontal resolution, and time step, data-driven ocean components can target different timescales and processes.

WenHai \cite{cui2025forecasting}, GLONET \cite{elaouni2025GLONET}, and FuXi-Ocean \cite{huang2025fuxiocean} use high-resolution GLORYS or HYCOM reanalysis products to produce ocean forecasts over approximately ten days. WenHai operates at $1/12^\circ$ resolution with a one-day time step and represents the upper 643~m. GLONET operates at $1/4^\circ$ resolution with a one-day time step and represents the full ocean depth. FuXi-Ocean operates at $1/4^\circ$ resolution with a six-hour time step and extends to approximately 1500~m. These models are evaluated against observations following the GODAE OceanView Intercomparison and Validation Task Team Class~4 framework for operational global ocean forecasting systems \cite{ryan2015godae}.

ORCA-DL targets monthly-to-decadal prediction using monthly time steps \cite{guo2025ORCA-DL}. It is pretrained on CMIP6 simulations and fine-tuned using the Simple Ocean Data Assimilation and Ocean Reanalysis System~5 datasets. ORCA-DL focuses on the upper 300~m, demonstrates skill at lead times of approximately one year, and shows preliminary skill on decadal timescales.

Samudra targets longer climate integrations by learning from GFDL OM4 simulations a 5-day autoregressive prognostic map, rather than numerically integrating OM4's governing equations at a 15-min timestep \cite{dheeshjith2024Samudra}. Given recent ocean states and prescribed atmospheric surface forcing, it evolves full-depth potential temperature, salinity, horizontal velocity, and sea-surface height, remaining stable in multicentury control-like integrations at approximately $1^\circ$ resolution. Samudra 2 combines a wider neural architecture with a dynamically reweighted loss to reduce the loss of temporal variability and the leakage of fast velocity patterns into slowly varying deep-ocean fields \cite{yuan2026samudra2scalingocean}. At $1^\circ$, it better captures upper-ocean variability and reduces deep-ocean temperature error; $1/2^\circ$ and $1/4^\circ$ versions support $\approx 8$-year rollouts and recover progressively finer western boundary currents and mesoscale eddies, although deep-ocean variability remains challenging.

\paragraph{Coupling atmosphere and ocean: Lessons from SamudrACE}

Data-driven models need not reproduce the component structure of conventional climate models. A single learned model can represent atmosphere--ocean memory implicitly, as in some subseasonal, seasonal, and decadal prediction systems. Alternatively, independently trained atmosphere and ocean emulators can be coupled through exchanged surface states and fluxes, following the modular structure of conventional climate models. The latter approach permits separate component development and makes the coupling variables explicit.

SamudrACE couples ACE2 for the atmosphere and land surface with Samudra for the ocean \cite{duncan2026samudrace,dheeshjith2024Samudra}. The atmospheric component uses a six-hour time step, while the ocean component uses a five-day time step. The system was trained to emulate a 200-year preindustrial simulation from GFDL CM4 \cite{Hel2019}. Samudra evolves potential temperature, salinity, velocity, and sea-surface height over the full ocean depth. Coupling required the addition of sea-ice variables and consistent land, ocean, and sea-ice fractional coverage across model components.

\begin{figure}
\centering
\includegraphics[width=\linewidth]{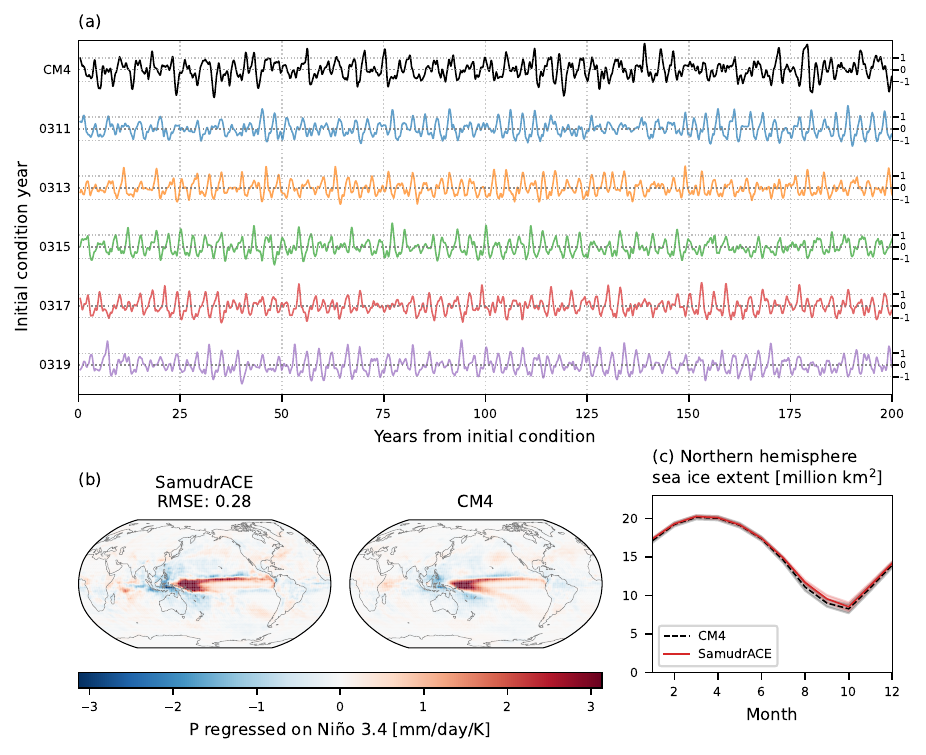}
\caption{\textbf{Coupled variability in SamudrACE.}
Five 200-year SamudrACE rollouts (colored lines) starting from initial conditions in the held-out period of the preindustrial CM4 simulation reproduce the approximate amplitude of Ni\~no3.4 variability (a).  Similarly, the ENSO precipitation response (b) and the seasonal sea-ice climatology (c) are approximately reproduced in a 40-year run during the hold-out period. However, SamudrACE exhibits a stronger preference for approximately three-year ENSO oscillations than CM4. Adapted from Figs.~3~and~4 of \cite{duncan2026samudrace}.}
\label{fig:SamudrACE}
\end{figure}

SamudrACE produces centuries-long simulations at approximately $1^\circ$ resolution and runs at roughly 1500 simulated years per wall-clock day on one NVIDIA H100 GPU. It reproduces CM4's mean climate and generates coupled variability that is absent when boundary conditions are prescribed. Its ENSO variability has a realistic amplitude, although it favors approximately three-year oscillations more strongly than the broader low-frequency spectrum in CM4. Its regression of precipitation on the Ni\~no3.4 index resembles that of CM4, and it reproduces seasonal sea-ice coverage in both hemispheres (Fig.~\ref{fig:SamudrACE}). In 200-year rollouts, the ocean component also represents multidecadal variability, including decadal variability in the Atlantic Meridional Overturning Circulation.

The next ongoing task is coupled emulation under changing external forcing. A near-term objective is to reproduce the CMIP DECK experiments---preindustrial control, historical, $1\%~\mathrm{yr}^{-1}$ CO$_2$ increase, and abrupt $4\times$CO$_2$---together with standard emissions scenarios. The aim is to achieve high fidelity to the reference coupled model using a limited training and validation dataset, for example a few hundred years of reference-model output. If successful, coupled emulators could generate large initial-condition ensembles at less than $1\%$ of the computational cost of the corresponding physical-model ensemble.

A further option is mixed AI--physics coupling: ACE2-NEMO couples a learned atmosphere emulator to a conventional ocean general circulation model \cite{antonio2026ace2nemocouplingmlatmospheric}. Such configurations fall under the broad umbrella of hybrid modeling strategies and could reduce the cost of component calibration. Their generalization under changing climates remains to be tested.

\section{What makes AI-powered climate models faster?\label{sec:SYPD}}

As AI permeates climate modeling, a practical question arises: when do AI-powered models reduce time to solution compared with dynamical models, and what enables that gain?

Time to solution, here defined as the wall-clock time required to complete a satisfactory simulation, limits ensemble size and scenario exploration when large. Yet reported performance is difficult to compare across model families because configurations differ in hardware, precision, scaling regime, prognostic state, effective resolution, and benchmarked tasks.

We use a deliberately simple comparison rather than an apples-to-apples benchmark, restricting dynamical models to GPU-ported configurations to make the comparison meaningful. Let $\mathrm{SYPD}$ denote simulated years per wall-clock day; $n_{\mathrm{acc}}$, $n_{\mathrm{prog}}$, and $n_{\mathrm{vert}}$ denote numbers of accelerators, reported prognostic variables, and vertical levels; and $\Delta x$ and $\Delta t$ denote nominal horizontal grid spacing and time step (reported interval between successive updates of the prognostic state). A normalized throughput proxy is
\begin{equation}
\mathrm{SYPD}_\mathrm{norm}=\frac{\mathrm{SYPD}\ n_{\mathrm{prog}} n_{\mathrm{vert}}}{n_{\mathrm{acc}} }.
\label{eq:normalized_throughput}
\end{equation}

For models advancing a full prognostic state, $\mathrm{SYPD}_{\mathrm{norm}}$ should increase approximately with $(\Delta x)^2\Delta t$. Note that AI models allow greater flexibility in the choice of $\Delta t$ as they need not explicitly resolve every numerical substep for stability.

\begin{figure}
\centering
\includegraphics[width=\linewidth]{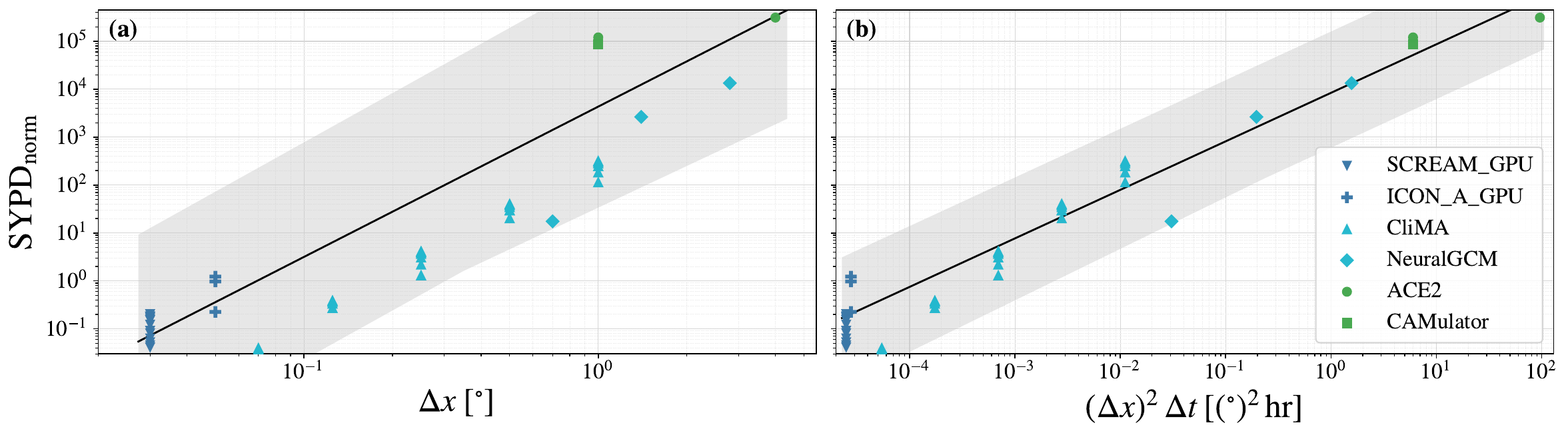}
\caption{\textbf{Reported normalized throughput across GPU-accelerated dynamical, hybrid AI--physics, and AI emulator models.} Each point represents one reported model configuration. Panel~(a) plots $\mathrm{SYPD}_{\mathrm{norm}}$ against nominal horizontal grid spacing $\Delta x$, whereas panel~(b) accounts for the reported prognostic update time-interval through $(\Delta x)^2\Delta t$; both panels share the same ordinate scale. Black lines are descriptive weighted-least-squares fits in log--log space, assigning equal total weight to each model. Gray envelopes are nominal 95\% prediction intervals.}
\label{fig:SYPD}
\end{figure}

Fig.~\ref{fig:SYPD} compares reported configurations of GPU-accelerated dynamical models (SCREAM \cite{donahue2024SCREAM} and ICON-A \cite{giorgetta2022ICONA_GPU}), hybrid AI--physics models (CliMA \cite{yatunin2026CliMA_dycore} and NeuralGCM \cite{kochkov2024neural}), and AI climate emulators (ACE2 \cite{watt2025ace2} and CAMulator \cite{chapman2025camulator}). Because throughput is conventionally reported as SYPD at a given horizontal resolution \cite{balaji2017CPMIP}, panel~(a) first fits $\mathrm{SYPD}_{\mathrm{norm}}$ in log--log space against $\Delta x$ alone, assigning equal total weight to each model. This comparison can give the impression that AI models are intrinsically more efficient, as they lie systematically above the dynamical models. Accounting for the interval between successive prognostic-state updates $\Delta t$ in panel~(b) largely removes this separation: the mean absolute model-level offset decreases from $0.81$ to $0.24$ in log-space, the weighted $R^2$ increases from $0.84$ to $0.96$, and the fitted exponent changes from $3.14$ with respect to $\Delta x$ to $1.01$ with respect to $(\Delta x)^2\Delta t$. The latter gives the approximate scaling $\mathrm{SYPD}_{\mathrm{norm}}\approx8.41\times10^3\left[(\Delta x)^2\Delta t\right]^{1.0}$. The gray bands show separately constructed, nominal 95\% model-balanced, leave-one-model-family-out prediction intervals for reported configurations.

Note that this descriptive fit is not a scaling law as the comparison spans different benchmarking standards as evidenced by the vertical superposition of data points for strong scaling tests. It also mixes mature, optimized GPU ports with mostly research AI inference implementations, and excludes AI training and training-data generation costs. Moreover, the high $R^2$ in panel~(b) partly reflects the seven-decade range of the data.

While our comparison cannot isolate intrinsic algorithmic efficiency, it shows why raw SYPD comparisons can mislead: Rather than escaping climate-model scaling altogether, AI methods are more readily GPU-ported and act directly on the main factors in Eq.~\ref{eq:normalized_throughput}. They can reduce $n_{\mathrm{prog}}$ by predicting only the variables needed for a task, reduce $n_{\mathrm{vert}}$ through compressed vertical representations, increase $\Delta t$ by not having the numerical stability constraints of dynamical models, or operate at coarser $\Delta x$ when fine-scale structure is not prognosed explicitly. Thus, much of AI's time-to-solution gain arises from computing less, and our proxy provides a rough estimate of how reported throughput varies with these choices.

This motivates downscaling: If we target a local hazard variable at high spatiotemporal resolution, the most efficient strategy may not be to prognose the full high-frequency, multivariate climate state at that resolution. A coarse global or regional model can instead provide the large-scale, scenario-dependent prognostic context, while a downscaling strategy such as GenFocal \cite{wan2026genFocal}, Google's R2-D2 \cite{lopezgomez2025dynamical_ML_downscaling} or HiRO-ACE \cite{perkins2025hiroace} reconstructs the high-resolution variables needed for hazard and eventually impacts modeling. In this sense, the strongest scaling advantage of AI may come less from faster full-state integration than from changing what is integrated in time, at what spatial resolution and time step, and what is reconstructed based on needs.

\section{AI-powered downscaling strategies}

\begin{figure}
\centering
\includegraphics[width=\linewidth]{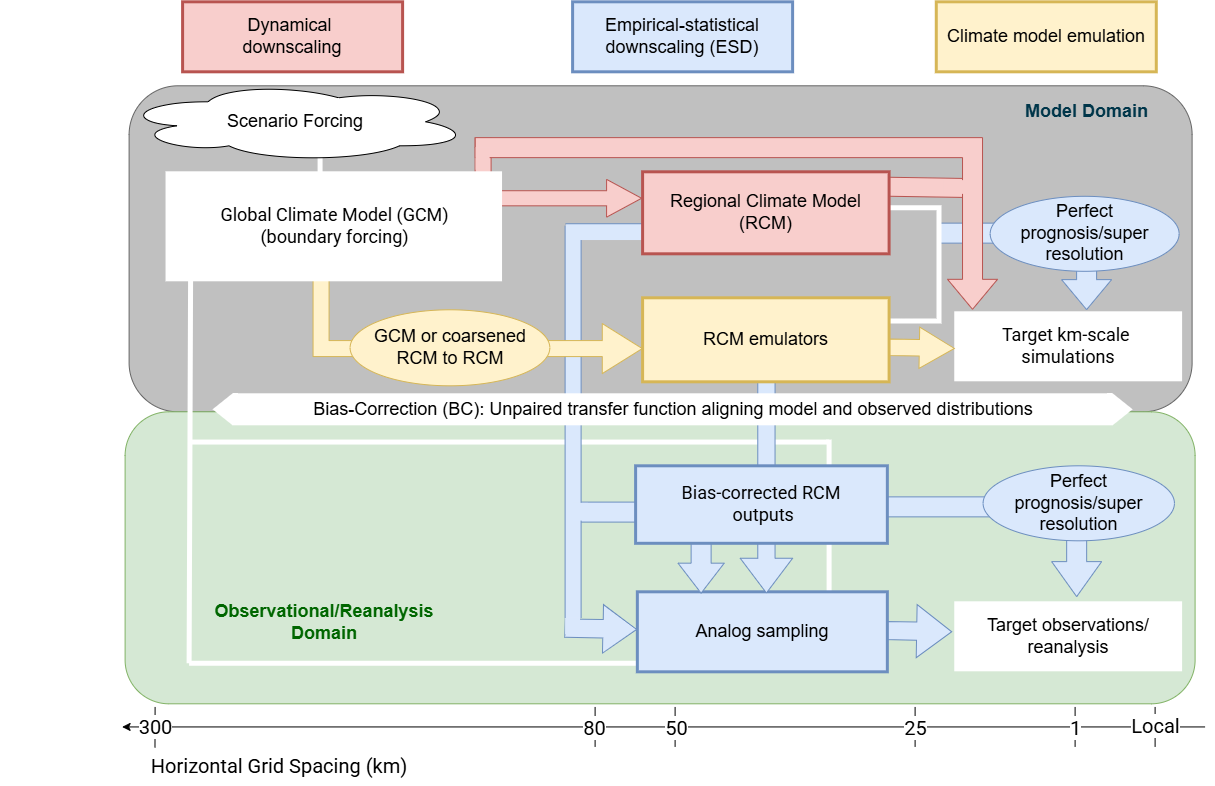}
\caption{\textbf{Modeling choices for downscaling.} Downscaling pathways connect scenario-forced global climate simulations to kilometer-scale simulations (center right) or observations and reanalyses (bottom right). The diagram distinguishes paired mappings, such as emulation and super-resolution, from unpaired bias correction across the model (top) and observational (bottom) domains; horizontal position indicates approximate horizontal grid spacing.}
\label{fig:downscaling}
\end{figure}

Ideally, local hazard assessment would draw directly on samples from the distribution of fields at high spatial and temporal resolution, such as kilometer-scale hourly precipitation, wind, or temperature, conditioned on a forcing scenario ${\cal S}$. On climate timescales, this requires long simulations and large ensembles, which remain computationally prohibitive for current storm-resolving models \cite{prein2026single,segura2025nextgems}. Despite emerging prototypes such as cBottle, which generates kilometer-scale atmospheric states conditioned on simplified forcings \cite{brenowitz2025cBottle}, most climate downscaling remains indirectly conditioned on ${\cal S}$ through coarse global climate-model trajectories $X_{\mathrm{GCM},{\cal S}}$:

\begin{equation}
p_{\mathrm{goal}}\left(X_{\mathrm{high\text{-}res}}\mid{\cal S}\right)\approx D_{\theta,\phi}\left(X_{\mathrm{high\text{-}res}}\mid X_{\mathrm{GCM},{\cal S}},Z\right),
\label{eq:downscaling_goal}
\end{equation}
where $p_{\mathrm{goal}}$ is the desired high-resolution climate distribution, $D_{\theta,\phi}$ is the conditional distribution induced by the downscaling procedure, and $Z$ denotes auxiliary predictors such as topography or land use. Starting from a scenario-forced GCM, downscaling can follow three broad routes (Fig.~\ref{fig:downscaling}). (1) Dynamical downscaling uses GCM boundary conditions to drive an RCM that explicitly simulates finer-scale regional processes \cite{giorgi2019thirty_years_RCMs}. (2) Empirical-statistical downscaling (ESD) uses data-driven algorithms to map GCM or RCM fields to higher-resolution observations, reanalyses, or, in some cases, reference-model output \cite{SUN2024review_statistical_downscaling_meteo}. Because observations provide no ground truth for future climates, observationally trained ESD methods must extrapolate beyond their targets and face OOD limitations, particularly for extremes \cite{rampal2024review_ML_regional_climate_downscaling}. (3) Climate-model emulation for downscaling instead learns from kilometer-scale regional simulations to approximate the mapping from a GCM, or from a coarsened RCM field, to high-resolution RCM output at lower computational cost, but depends on scarce fine-scale simulation targets \cite{addison2026machine}. When simulation archives span multiple scenarios, such emulators can be trained and evaluated across climates and rapidly provide local climate projections \cite{kendon2025potential}. These routes are not mutually exclusive: statistical refinement can follow partial dynamical downscaling \cite{walton2015hybrid_stat_dyn}.

Recent advances in AI-powered downscaling can be organized around three themes: (1) generative modeling, (2) distributional and process-oriented evaluation, and (3) unpaired bias correction.

First, super-resolution illustrates the need for generative modeling. Mapping a coarse field to a corresponding high-resolution field is underdetermined: the coarse field does not uniquely determine the small-scale structures that could have produced it. A deterministic model can estimate a conditional mean or another central tendency, but it cannot represent the range of high-resolution fields compatible with the same coarse input. This limitation is especially important for precipitation, near-surface winds, and other intermittent or spatially heterogeneous variables. Modern super-resolution for climate applications therefore increasingly targets a distribution of high-resolution fields conditioned on low-resolution fields rather than a deterministic reconstruction:

\begin{equation}
X_{\mathrm{high\text{-}res}}\sim p_{\theta}\left(
X_{\mathrm{high\text{-}res}}\mid X_{\mathrm{low\text{-}res}},Z
\right),
\qquad
{\cal C}X_{\mathrm{high\text{-}res}}=X_{\mathrm{low\text{-}res}},
\label{eq:paired_downscaling}
\end{equation}

where $Z$ denotes auxiliary predictors and ${\cal C}$ is a coarsening operator. In paired super-resolution of the same variable, the generated high-resolution field is typically required to remain consistent with the coarse field it refines. Hard-constrained downscaling enforces this consistency in the model architecture \cite{harder2023hard_constrained_DL_downscaling}. More generally, the coarse predictor and high-resolution target need not be the same variable; such paired mappings belong to the broader ``perfect prognosis'' family of statistical downscaling methods \cite{Maraun_Widmann_2018}. For example, coarse moisture may be more predictive of high-resolution precipitation than coarse precipitation itself.

This generative view has changed model design. Diffusion and flow-based models are now prominent in downscaling because they can recover fine-scale spatial and spectral structure while representing multiple plausible outcomes \cite{watt2024generativediffusionbaseddownscalingclimate}. Corrective Diffusion (CorrDiff) separates a deterministic prediction from a residual diffusion model that generates fine-scale variability not determined by the coarse input \cite{mardani2025residual}. Video diffusion extends generative downscaling to spatiotemporal precipitation sequences and can support super-resolution in both space and time \cite{srivastava2024precip_downscaling_video_diffusion,defez2026scaleadaptiveSR}. Flow matching instead learns a velocity field that transports a reference distribution, such as that of encoded coarse fields, toward the conditional target distribution of e.g., high-resolution fields \cite{fotiadis2025adaptive,debeire2026physics}. The same generative logic applies to climate-model emulation, where the target is the conditional variability of a high-resolution dynamical model rather than super-resolved fields \cite{schillinger2026enscale}.

Second, once downscaling is treated generatively, evaluation must go beyond mean error. Recent benchmarks such as CORDEX-ML-Bench show that some AI models can perform comparatively poorly on climatological metrics despite excelling on metrics standard in computer vision \cite{rampal2026cordex}. Consequently, evaluation frameworks now combine computer-vision metrics with tools from statistical downscaling and probabilistic forecasting. Proper scores require both calibration and sharpness of the conditional distribution, while multivariate distributional distances assess whether variables are realistic jointly rather than only marginally. Extreme-value diagnostics assess whether distribution tails are well calibrated, including return levels, high quantiles, and application-derived compound diagnostics such as wildfire risk \cite{lopezgomez2025dynamical_ML_downscaling}. In physical space, object-based metrics assess whether convective cells, fronts, rainbands, or cyclone structures have realistic size, shape, and organization, with recent work approximating these often discontinuous metrics by differentiable surrogates for optimization \cite{quarenghi2026minkowski}. Temporal-coherence metrics test whether generated fields flicker between frames or evolve consistently, while spectral metrics test whether variance is present at the targeted spatial scales, because nominal grid spacing alone can be misleading. Benchmarks increasingly test transfer across regions \cite{harder2026benchmarking}, variables, and data products \cite{prasad2024evaluating}, adding OOD generalization as a further evaluation requirement. No single metric is sufficient, so diagnostics should be selected for the variable, scale, and application of interest \cite{maraun2015value}.

Third, climate applications sharpen the distinction between paired super-resolution and unpaired bias correction. Paired super-resolution assumes sample-wise correspondence between low- and high-resolution fields, as when the low-resolution input is obtained by coarsening the target (Eq.~\ref{eq:paired_downscaling}). A coupled-ESM trajectory, however, is not aligned frame by frame with observations, even over the same historical period, because internal variability and model biases make the two realizations diverge. To reconcile model output with observations statistically, the goal is to transform the output $X_M$ of model $M$ so that its corrected distribution matches the observed distribution conditional on the relevant coarse information:

\begin{equation}
{\cal T}_{\phi}^{M}\left(X_M\right)\,\big|\, X_{\mathrm{coarse}},Z\ \overset{d}{\approx}\ X_{\mathrm{obs}}\,\big|\, X_{\mathrm{coarse}},Z,\qquad M\in\left\{\mathrm{GCM},\mathrm{RCM}\circ\mathrm{GCM},\ldots\right\},
\label{eq:unpaired_bias_correction}
\end{equation}

where ${\cal T}_{\phi}^{M}$ is a model-specific transfer function and $\overset{d}{\approx}$ denotes approximate equality in distribution. Conditioning on the coarse state and auxiliary predictors makes this criterion stricter than matching climatological marginals but more realistic than requiring sample-wise alignment.

Statistical bias correction addressed this problem before the advent of AI-powered downscaling. Multivariate methods move beyond univariate quantile mapping by adjusting dependence across variables or locations \cite{francois2020multivariateBC}, including through multivariate quantile mapping \cite{cannon2018multivariate}, optimal transport \cite{robin2019BC_OT}, and rank resampling \cite{vrac2018R2D2}. Ensemble corrections use members jointly to correct the target distribution while preserving internal variability \cite{vaittinada2021ensemble}. AI extends these approaches by parameterizing higher-dimensional transfer functions that act on joint spatial, temporal, and multivariate distributions rather than isolated grid points. This motivates a two-stage pipeline: unpaired, model-specific bias correction through ${\cal T}_{\phi}^{M}$, followed by sampling from $p_{\theta}$ for paired spatiotemporal super-resolution: 
$D_{\theta,\phi}\!\left(X_{\mathrm{high\text{-}res}}\mid X_{M},Z\right)=p_{\theta}\!\left(X_{\mathrm{high\text{-}res}}\ \middle|\ {\cal T}_{\phi}^{M}\!\left(X_{M}\right),Z\right)$.

Methods that neglect coarse-predictor bias, or attempt single-step downscaling via super-resolution, can fail when the coarse predictor distribution differs from the observational target distribution \cite{reddy2025limitation}, leading to artifacts when used with future GCM conditions \cite{gonzales-abad2025DL_climate_projections}.

Recent models therefore make the paired--unpaired distinction more explicit. Aich et al. \cite{aich2026conditionalBC} use a shared latent space to align observational and ESM precipitation fields before training a conditional diffusion model to simultaneously downscale and correct the ESM. GenFocal frames downscaling as high-dimensional distribution matching, separating coarse-scale generative bias correction from super-resolution \cite{wan2026genFocal}. Unpaired AI bias correction nevertheless remains less established than paired AI super-resolution, and its advantage over established statistical methods is not yet clear-cut: statistical correction can better reproduce local marginals, AI can better recover spatial structure, and combining the two can outperform either alone \cite{hess2023deep}. Establishing AI's added value therefore requires comparing joint distributional fidelity, forced-trend preservation, and whether the compute and data requirements justify the skill against strong statistical baselines, ideally across multiple GCM/RCM chains.

For climate applications, the final test is robustness, which can be formalized through well-defined generalization errors \cite{largeau2026robustnessPextremes_climate}. A downscaling method may perform well in the present climate yet fail for rare extremes absent from training, altered circulation patterns, or future thermodynamic states. Pseudo-reality experiments, also called model-as-truth or perfect-model experiments, provide one way to test this failure mode: high-resolution model output is treated as pseudo-observations, and the downscaler is evaluated on scenarios or extreme values withheld during training \cite{rampal2024extrapolation_GAN}. Robustness should be assessed for the full pipeline, including bias correction and super-resolution, using complementary diagnostics of spectra, temporal coherence, intervariable consistency, and extremes. This is especially important for hazards whose impacts depend on coherent spatiotemporal evolution, such as tropical cyclones, heatwaves, and compound extremes.

A central open question is how much, and what range of, climate data are needed for reliable extrapolation. CORDEX-ML-Bench shows that AI downscalers trained only on historical conditions can underestimate future climate-change signals, whereas including warmer climates in training improves their representation \cite{rampal2026cordex}. Benchmarks should therefore evaluate complete pipelines under withheld climates and extremes, with particular attention to forced-trend preservation and compute-to-skill tradeoffs, which can differ substantially across generative methods \cite{ward2025intercomparison}. Establishing this robustness is essential for decision-making because inexpensive downscaling can generate the large ensembles needed to quantify fine-scale internal variability and the probabilities of rare or potentially unprecedented local hazards \cite{rampal2026downscaling,lewis2025generative}.

\section{Conclusion}
\subsection{Summary}

AI is already changing operational weather forecasting, but extending these gains to climate-relevant horizons requires more than longer rollouts of weather models. The central challenge is not only to maintain forecast skill, but to produce credible responses to altered forcings, unseen extremes, and counterfactual scenarios. This is not simply a shift from initial-condition to boundary-condition prediction, but a fundamental change in both the modeling problem and the evaluation criteria. AI has accelerated the development of deterministic and probabilistic weather forecasters, subseasonal prediction systems, full climate model emulators, post-processing frameworks, and AI-based downscaling, but none of these systems in isolation resolves the full problem of Earth system prediction across timescales.

To address the OOD nature of climate change prediction, combining AI's ability to learn from data with the physical structure needed for robust OOD prediction has emerged repeatedly throughout this review, with two main requirements. First, forcing agents such as greenhouse gases, aerosols, land-use change, and more generally perturbations of the hydrological and biogeochemical cycles must be represented explicitly enough for independent manipulation. Second, models must be evaluated for robustness beyond their training distribution, including rare extremes, transient adjustments, and climate shifts. These requirements favor modular combinations of data-driven and physics-based components, with the optimal mix depending on the application of interest.

Within this broader framework, AI can improve Earth system modeling in four main ways. First, AI can enhance model physics through learned parameterizations of unresolved processes such as clouds and turbulence, trained on observations and high-resolution simulations but embedded within models that retain physical consistency and conservation laws; distilling AI parameterizations additionally promotes process understanding. Second, AI can accelerate discovery through climate model emulators, which approximate expensive models at far lower cost, enabling larger ensembles to explore parameter and scenario uncertainties represented in the training data and making advanced climate information more accessible for scientific and downstream use. Third, AI can make climate projections actionable through statistical downscaling, including super-resolution and hazard-specific risk modeling. Generative models can translate coarse 50--200~km climate information into kilometer-scale fields more efficiently than dynamical downscaling, while combinations of climate output with topography, vegetation, or the built environment can support targeted estimates of hazards such as storm surge and wildfire, improving the associated risk estimation. Fourth, AI can synthesize information across the heterogeneous data streams on which climate science depends, fusing models, satellites, and in-situ observations into products of higher value than any single source, e.g., by combining microwave, laser, and lidar measurements to better constrain liquid and ice water path, which are difficult to observe reliably from a single source.

These roles are synergistic. Better fused and downscaled data provide stronger targets for learned parameterizations, improved physics increases the realism and value of fast emulators, and all three improve the quality of actionable climate information. As comparing AI against optimized GPU-ported models highlights that AI reduces the time-to-solution by computing less (timesteps, prognostic variables, levels, etc.), its greatest promise may therefore lie in complementing and reorganizing the use of physics-based models through the accelerated model development cycle AI allows. This naturally points toward an Earth system modeling enterprise built from transparent combinations of learned and physically-based components, with different models specializing by regime, process, task, and scale, with the caveat that this may complicate the AI design and training process required to obtain optimal results. Coding the physically-based components in an ML-differentiable language is key for optimizing this training across the full modeling system \cite{Gelbrecht2023differentiableESM}.

\subsection{Outlook}

\subsubsection{ERA6 and CMIP7 as near-term drivers of progress}

Near-term progress in Earth system AI may be driven by improved weather and climate data infrastructures, especially around ERA6 and CMIP7. On the weather side, improved reanalyses, higher spatial resolution, better stratospheric representation, stronger ocean--atmosphere coupling \cite{hahner2026representingsurfaceoceanecmwfs}, and continued advances in physics and data assimilation should strengthen the training signal available to AI systems and improve their handling of extremes. On the climate side, CMIP7-related efforts will expand the space of scenarios, attribution experiments, and likely emissions-driven simulations, providing a richer basis for learning and testing forced responses, counterfactual climates, and process-level interactions involving the four CMIP7 challenges: SST patterns, weather extremes, carbon--water coupling, and possible tipping behavior.

\subsubsection{Towards an ecosystem of transparent, hybrid prediction systems}

Given that AI is democratizing modeling, the future of Earth system AI may not be dominated by a single foundation model, but rather evolve towards an ecosystem of specialized, mixable, complementary models whose domains of validity would ideally be explicit. On the weather side, models may specialize by region, dynamical regime, loss function, or observing system, for example separating components optimized for bulk skill from those optimized for extremes, or remote-sensing-informed models from reanalysis-informed ones. On the climate side, models may specialize by Earth system process, diagnostic task, bias-correction role, domain of validity, or data source, e.g., distinguishing components trained on fused observations from those trained as emulators of physically-based simulations. A successful example is the parameterization mixing framework recently introduced in the ICON-A-MLe hybrid model, where a conventional convection parameterization is mixed with an AI parameterization based on its learned confidence \cite{heuer2025confidenceHybridICON}, also demonstrating inter-model transferability as the AI parameterization was trained on E3SM data. 

Model ecosystems support ensembles by allowing uncertainty to arise not only from generative modeling or stochastic sampling of weights and training data, but also from structural diversity across models. They support data fusion by allowing models trained on different observational products and reference simulations to contribute where they are most informative. They support hazard-specific prediction through task-aware models for extremes and impact-related variables. They also create room for creativity in model design, because Earth system AI need not inherit the structure of legacy forecast systems if more direct formulations and decompositions better match the target process, the available data, or the prediction horizon.

Such diversity increases the burden of evaluation and maintenance, and credibility may therefore concentrate around a smaller number of well-maintained ``anchor models,'' while more specialized models address narrower tasks. At the same time, AI climate emulators and hybrid AI--physics models may converge toward a broader ecosystem of transparent hybrid systems, combining learned and physically based components differently across applications. For such an ecosystem to remain credible, each component should ideally declare the data on which it was trained, its optimization objective, its sample-wise skill and uncertainty, and its training and inference cost. This is not only a practical requirement for benchmarking; it is also a sustainability requirement for deciding whether a model's added value is worth the negative environmental impact associated with training and inference, and a scientific requirement for understanding a model's trustworthiness. The central tension is therefore between the proliferation coming from lower development barriers and the consolidation around models that are flexible, transparent, and trusted. If managed successfully, this ecosystem could combine dynamical understanding, observations, simulations, and task-specific learning into more adaptive and testable prediction systems.

\section*{Acknowledgments}
\paragraph*{Funding:} 
TB and SA acknowledge support from the Swiss National Science Foundation (SNSF) under Grant No. 10001754 (“RobustSR” project). TB also acknowledges support from AIPEX, funded by the Swiss State Secretariat for Education, Research and Innovation (SERI, Grant 23.00546). JDN acknowledges support from the R. P. \& L.S. Turco Chair and from U.S. National Science Foundation (NSF) grant AGS-2414576.  HS acknowledges the support from the Hong Kong Jockey Club Charities Trust (FA123), and the Innovation and Technology Commission projects (ITP/047/23LP and ITS/003/24SC), and the State Key Laboratory scheme (ITC-SKLCRCC26EG01). TS and CC were supported by Schmidt Sciences, LLC, The U.S.\ Office of Naval Research (grant N00014-23-1-2654), and the U.S.\ NSF (grant 2530746). CB, SKC, and OWM acknowledge the Allen Family Fund for Science and Technology for its generous support of Ai2.

\paragraph*{Author contributions:}
Conceptualization: T.B., J.D.N., H.S., C.B., C.C., I.L.-G., T.S., O.W.-M. Formal analysis \& Investigation: T.B. Visualization: T.B., J.D.N., H.S., S.A., C.B., W.C., C.C., S.K.C., T.S., A.S., O.W.-M. Writing – original draft: T.B., J.D.N., H.S., C.B., T.S., O.W.-M. Writing – review \& editing: T.B., J.D.N., H.S., S.A., C.B., W.C., C.C., S.K.C., A.G., I.L.-G., T.S., A.S., O.W.-M.
\paragraph*{Competing interests:}
There are no competing interests to declare.

\paragraph*{Data and materials availability:}
Data are available associated with the publications reviewed. Data required to reproduce Figures \ref{fig:Saola_Forecast} and \ref{fig:SYPD} are available on Zenodo (DOI: 10.5281/zenodo.22872186), and so is the corresponding analysis code (DOI: 10.5281/zenodo.22873425). Processed data and analysis code required to reproduce Figure \ref{fig:clima-s2s} are also available on Zenodo (DOI: 10.5281/zenodo.22681289).

\paragraph*{AI use declaration:}
We used large language models (LLMs) for language editing only, specifically to assist in converting point form notes from multi-author discussions to paragraphs form, in reducing manuscript length from the original draft and in improving syntax and readability. No scientific data or figure was generated by LLMs.

\bibliographystyle{unsrt}  
\bibliography{references}

\end{document}